\documentclass[12pt]{article}
\usepackage{epsfig}
\usepackage{graphicx}
\usepackage{amsmath}

\begin{document}

 \newcommand{\beq}{\begin{equation}}
\newcommand{\eeq}{\end{equation}}
\newcommand{\bea}{\begin{eqnarray}}
\newcommand{\eea}{\end{eqnarray}}
\newcommand{\beqn}{\begin{eqnarray}}
\newcommand{\eeqn}{\end{eqnarray}}
\newcommand{\beas}{\begin{eqnarray*}}
\newcommand{\eeas}{\end{eqnarray*}}
\newcommand{\defi}{\stackrel{\rm def}{=}}
\newcommand{\non}{\nonumber}
\newcommand{\bquo}{\begin{quote}}
\newcommand{\enqu}{\end{quote}}
\newcommand{\qt}{\tilde q}
\newcommand{\m}{\tilde m}
\newcommand{\trho}{\tilde{\rho}}
\newcommand{\tn}{\tilde{n}}
\newcommand{\tN}{\tilde N}
\newcommand{\gsim}{\lower.7ex\hbox{$\;\stackrel{\textstyle>}{\sim}\;$}}
\newcommand{\lsim}{\lower.7ex\hbox{$\;\stackrel{\textstyle<}{\sim}\;$}}


\def\de{\partial}
\def\Tr{ \hbox{\rm Tr}}
\def\const{\hbox {\rm const.}}
\def\o{\over}
\def\im{\hbox{\rm Im}}
\def\re{\hbox{\rm Re}}
\def\bra{\langle}\def\ket{\rangle}
\def\Arg{\hbox {\rm Arg}}
\def\Re{\hbox {\rm Re}}
\def\Im{\hbox {\rm Im}}
\def\diag{\hbox{\rm diag}}


\def\QATOPD#1#2#3#4{{#3 \atopwithdelims#1#2 #4}}
\def\stackunder#1#2{\mathrel{\mathop{#2}\limits_{#1}}}
\def\stackreb#1#2{\mathrel{\mathop{#2}\limits_{#1}}}
\def\Tr{{\rm Tr}}
\def\res{{\rm res}}
\def\Bf#1{\mbox{\boldmath $#1$}}
\def\balpha{{\Bf\alpha}}
\def\bbeta{{\Bf\beta}}
\def\bgamma{{\Bf\gamma}}
\def\bnu{{\Bf\nu}}
\def\bmu{{\Bf\mu}}
\def\bphi{{\Bf\phi}}
\def\bPhi{{\Bf\Phi}}
\def\bomega{{\Bf\omega}}
\def\blambda{{\Bf\lambda}}
\def\brho{{\Bf\rho}}
\def\bsigma{{\bfit\sigma}}
\def\bxi{{\Bf\xi}}
\def\bbeta{{\Bf\eta}}
\def\d{\partial}
\def\der#1#2{\frac{\d{#1}}{\d{#2}}}
\def\Im{{\rm Im}}
\def\Re{{\rm Re}}
\def\rank{{\rm rank}}
\def\diag{{\rm diag}}
\def\2{{1\over 2}}
\def\ntwo{${\mathcal N}=2\;$}
\def\nfour{${\mathcal N}=4\;$}
\def\none{${\mathcal N}=1\;$}
\def\ntwot{${\mathcal N}=(2,2)\;$}
\def\ntwoo{${\mathcal N}=(0,2)\;$}
\def\x{\stackrel{\otimes}{,}}

\def\ba{\beq\new\begin{array}{c}}
\def\ea{\end{array}\eeq}
\def\be{\ba}
\def\ee{\ea}
\def\stackreb#1#2{\mathrel{\mathop{#2}\limits_{#1}}}

\def\Tr{{\rm Tr}}
\newcommand{\cpn}{CP$(N-1)\;$}
\newcommand{\wcpn}{wCP$_{N,\tilde{N}}(N_f-1)\;$}
\newcommand{\wcpd}{wCP$_{\tilde{N},N}(N_f-1)\;$}
\newcommand{\vp}{\varphi}
\newcommand{\pt}{\partial}
\newcommand{\ve}{\varepsilon}
\renewcommand{\theequation}{\thesection.\arabic{equation}}

\setcounter{footnote}0

\vfill

\begin{titlepage}

\begin{flushright}
FTPI-MINN-11/05, UMN-TH-2938/11\\
16 March, 2011
\end{flushright}

\vspace{2mm}

\begin{center}
{  \Large \bf  
Non-Abelian Duality and   
\\[0.5mm] 
Confinement: from \boldmath{\ntwo} to \boldmath{\none} 
\\[2,9mm]
Supersymmetric QCD
}

\vspace{3mm}

 {\large \bf    M.~Shifman$^{\,a}$ and \bf A.~Yung$^{\,\,a,b}$}
\end {center}

\begin{center}

$^a${\it  William I. Fine Theoretical Physics Institute,
University of Minnesota,
Minneapolis, MN 55455, USA}\\
$^{b}${\it Petersburg Nuclear Physics Institute, Gatchina, St. Petersburg
188300, Russia
}
\end{center}


\begin{center}
{\large\bf Abstract}
\end{center}

Recently we discovered and discussed non-Abelian duality in the {\em quark vacua} of \ntwo super-Yang--Mills 
theory with the U$(N)$ gauge group and $N_f$ flavors ($N_f>N$). Both theories from the dual 
pair support non-Abelian strings which confine monopoles. Now we introduce an \ntwo-breaking deformation, 
a mass term $\mu{\mathcal A}^2$
for the adjoint fields. Starting from a small deformation we eventually make it large which enforces 
complete decoupling of the adjoint fields. We show that the above non-Abelian duality fully survives in 
the limit of \none SQCD, albeit some technicalities change. For instance, non-Abelian strings which used 
to be BPS-saturated in the \ntwo limit, cease to be saturated in \none SQCD. Our 
duality is a distant relative of Seiberg's duality in \none SQCD. Both share some common features 
but have many drastic distinctions. This is due to the fact that Seiberg's duality apply to the monopole
rather than quark vacua.

 More specifically, in our theory we deal with   $N< N_f<\frac32 N $ {\em massive} quark flavors. 
 We  consider the vacuum in which  $N$ squarks condense.
Then we identify a crossover transition 
from weak to strong coupling. At strong coupling we find a dual theory, U$(N_f-N)$ SQCD, with   
$N_f$ light dyon flavors.
Dyons condense triggering the formation of non-Abelian strings which confine monopoles. 
Screened quarks and gauge bosons of the original theory
decay into confined monopole-antimonopole pairs  and form stringy mesons.

\vspace{2cm}

\end{titlepage}

 \newpage



\section {Introduction and setting the goal:  from \boldmath{\ntwo} to \boldmath{\none}}
\label{intro}
\setcounter{equation}{0}

The dual Meissner effect as the confinement mechanism   \cite{mandelstam}
in Yang--Mills theories remains obscure despite a remarkable breakthrough in \ntwo super\-symmetric 
theories,  where the exact Seiberg--Witten solution was found \cite{SW1,SW2}.
 Seiberg and Witten demonstrated \cite{SW1,SW2} that magnetic monopoles do condense
 in the so-called monopole vacua of the \ntwo theory after one switches on a small \ntwo-breaking deformation
 of the $\mu{\mathcal A}^2$ type. Upon condensation of the monopoles,  
 chromoelectric flux tubes (strings)  of the Abrikosov--Nielsen--Olesen
(ANO) type  \cite{ANO} are formed. This  leads to confinement of (probe) quarks  attached to
the endpoints of confining strings. 

 The Seiberg--Witten mechanism of confinement is essentially Abelian\,\footnote{By  non-Abelian
confinement we mean such dynamical regime in which at distances
of the flux tube formation all gauge bosons are equally important,
while the Abelian confinement occurs when the relevant gauge dynamics at such distances is Abelian.
Note that Abelian confinement can take place in non-Abelian theories,
the Seiberg--Witten solution is just one example.}
\cite{DS,HSZ,Strassler,VY,Yrev}. This is due to the fact that in the Seiberg--Witten solution
 the non-Abelian gauge group of the
underlying \ntwo theory (say,  SU$(N)$) is broken down to the Abelian subgroup U(1)$^{N-1}$  by
condensation (in the strongly coupled monopole vacua) of the {\em adjoint scalars} inherent to \ntwo.
The subsequent monopole condensation occurs essentially in the Abelian U(1)$^{N-1}$ theory.
This feature makes the \ntwo theory dissimilar from pure Yang--Mills, in which there is no dynamical 
Abelianization. Hence, to get closer to reality, a natural desire arises to eliminate the adjoint scalars,
passing if not to ${\mathcal N}=0$, at least to \none. That's what we will eventually do.

\begin{figure}
\epsfxsize=6cm
\centerline{\epsfbox{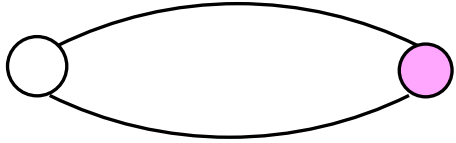}}
\caption{\small Meson formed by monopole-antimonopole pair connected by two strings.
Open and closed circles denote the monopole and antimonopole, respectively.}
\label{figmeson}
\end{figure}

However, \none theories do not
support monopoles (dyons), at least at the quasiclassical level, and the very meaning of the 
dual Meissner effect gets obscure.
In  search of a non-Abelian confinement mechanism similar in spirit to the  
 Meissner mechanism of Nambu, 't Hooft, and Mandelstam 
 we recently explored a different, albeit related,  scenario \cite{SYdual,SYtorkink}. 
To begin with, we focused on  the quark (rather than monopole) vacuum of 
\ntwo supersymmetric QCD (SQCD) 
with the U($N$) gauge group (rather than SU($N$)) and
$N_f$ flavors of fundamental quark hypermultiplets, with $N_f$ in the interval $N<N_f<2 N$. 
Then, there  is no confinement of the chromoelectric charges; on the contrary, they
are Higgs-screened. Instead, the chromomagnetic charges are confined by 
{\em non-Abelian} strings. 
They --- the chromomagnetic charges --- manifest themselves 
in a clear-cut manner as junctions of two  nonidentical, albeit degenerate,
strings. Moreover, at strong coupling  (where, as we will see, a  dual description is applicable) 
the quarks and
gauge bosons of the original theory  decay into monopole-antimonopole 
pairs on the curves of marginal stability (CMS).
The (anti)monopoles forming the pair are confined. In other words, the original quarks and gauge bosons 
evolve in the 
strong coupling domain  into ``stringy mesons" with two constituents 
being connected by two strings as shown in  Fig.~\ref{figmeson}.  
These mesons are expected to lie on Regge trajectories. 

All these phenomena take place in the quark vacua
of the \ntwo theory \cite{SYdual,SYtorkink}. This theory is slightly different from 
the Seiberg--Witten model.
Namely, as was mentioned, the U(1) gauge factor is added, and, then, the
Fayet--Iliopoulos (FI) \cite{FI} $D$-term $\xi$ 
is introduced. Then, we single out 
the vacuum in which  $r=N$ (s)quarks condense.  
A global color-flavor locked symmetry survives in the limit of equal quark mass terms.
At large $\xi$ this theory is at weak coupling and 
supports non-Abelian flux tubes (strings)
 \cite{HT1,ABEKY,SYmon,HT2} (see also \cite{Trev,Jrev,SYrev,Trev2} for reviews). It is the
formation of these strings that ensures confinement of monopoles. Upon reducing the FI parameter $\xi$, the theory
goes through a crossover transition \cite{SYdual,SYcross,SYcrossp} into a strongly 
coupled regime which can be described
in terms of weakly coupled {\em dual} \ntwo SQCD,
 with the U$(\tN)\times$U(1)$^{N-\tN}$ gauge group and $N_f$ flavors of
light {\em dyons}.\footnote{This is in perfect agreement with the
results obtained in \cite{APS} where the SU$(\tN)$ dual  gauge group  was identified
at the root of the baryonic Higgs branch in the  SU($N$) gauge theory with massless (s)quarks.}
Here
\beq
\tN=N_f-N\,,
\label{tN}
\eeq
as in Seiberg's duality in \none theories \cite{Sdual,IS}
where the emergence of the dual SU$(\tN)$ group was first observed.

The dual theory supports non-Abelian strings due to condensation of light dyons much in the same way
as the string formation in the original theory is due to condensation of quarks. Moreover, the number of 
distinct strings is, of course, the same in the original and dual theories.
The strings  of the dual theory confine monopoles too, 
rather than quarks \cite{SYdual}.
This is due to the fact that the  light dyons   condensing in the dual theory  
carry weight-like chromoelectric charges (in addition to chromomagnetic charges). In other words, they carry
the quark charges. The strings formed through condensation of these dyons
can confine only  states with the root-like magnetic charges, i.e. 
the monopoles  \cite{SYdual}.  Thus, 
our \ntwo non-Abelian duality is {\em not}  electromagnetic. 

 The chromoelectric charges of quarks (or gauge bosons)
are Higgs-screened a large $\xi$. As was mentioned above, 
in the domain of small $\xi$ (where the dual description is applicable) these states   
decay into the monopole-antimonopole  pairs on CMS,
see \cite{SYtorkink} for the proof of this fact.
The monopoles and antimonopoles forming the pair
 cannot abandon
each other  because they are confined. Therefore, the original quarks and gauge bosons, with the passage 
to the strong coupling domain of small $\xi$, evolve   into ``stringy mesons''  shown in  
Fig.~\ref{figmeson}. A detailed discussion of these stringy mesons can be found 
in  \cite{SYrev}. The same picture takes place when we move in the opposite direction,
with the interchange of two theories from the dual pair.

Deep in the non-Abelian quantum regime the confined monopoles were demonstrated \cite{SYtorkink} 
to belong to the {\em fundamental representation} of the global  (color-flavor locked) group. 
Therefore, the  monopole-antimonopole
mesons can be both, in the adjoint and singlet representation of this  group. 
This  pattern of confinement seems to be similar to what we have in actuality, except that
the role of the ``constituent quarks'' inside mesons is played by the monopoles.

Low-energy dynamics on the world sheet of
the non-Abelian strings under discussion
are described by two-dimensional CP models \cite{HT1,ABEKY,SYmon,HT2}. 
From the world-sheet standpoint different (degenerate) non-Abelian strings 
are different supersymmetric vacua of the CP models.
Confined monopoles are in fact  kinks interpolating between these vacua. Nonperturbative generation of 
the dynamical scale $\Lambda_{\rm CP}$  stabilizes the  kink inverse sizes and masses at
 $O(\Lambda_{\rm CP})$ \cite{SYmon,SYrev}. This is in contradistinction with the 
absence of {\em classical} stabilization of monopoles in the non-Abelian regime (see e.g the discussion of  the so-called ``monopole clouds'' in  \cite{We}).\footnote{To better explain this statement we should point out that,
say, in the monopole vacua of \ntwo SQCD,  the Seiberg--Witten solution tells us \cite{SW1,SW2} that
the theory 
dynamically Abelianizes. That's why it is so difficult to apply the standard confinement scenario,
based on monopole condensation,  to theories without adjoint scalars.
We just do not know what does that mean,  non-Abelian monopole in the Higgs/Coulomb phase.}

In this paper we report on the second stage of the program, namely the study of non-Abelian 
duality in the {\em absence} of the adjoint fields, in \none SQCD. To pass from \ntwo to \none we add a deformation term
$\mu{\mathcal A}^2$ in the superpotential.
We
show that the picture of the non-Abelian monopole confinement outlined above for \ntwo
survives this deformation all the way up to large $\mu$ where
the adjoint fields decouple leaving us with \none SQCD.

We start  our work from  \ntwo SQCD with the  U($N$) gauge group
and $N_f$  {\em  massive} quark flavors where
\beq
N< N_f<\frac32 N
\label{irfr}
\eeq
to ensure infrared freedom in the dual theory at large $\mu$.
Since the deformation superpotential (\ref{msuperpotbr}) plays the role of an effective FI 
term (being combined with the nonvanishing quark mass terms), there is no need to introduce the FI 
term through the $D$ term. Although it is certainly 
doable, this would be nothing but an unnecessary complication.

At small $\mu$ the deformation superpotential (\ref{msuperpotbr}) reduces to the
Fayet--Iliopoulos $F$-term with the effective FI parameter $\xi$
determined by $\xi\sim \sqrt{\mu m}$, where $m$ presents a typical
scale of the quark masses. 
We focus on the so-called $r$  vacuum in which $r=N$ quarks condense, with the subsequent formation 
of the non-Abelian strings which confine  monopoles. Much in the same way as in our previous \ntwo studies
\cite{SYdual,SYtorkink}  with the Fayet--Iliopoulos  $D$-term,  there is a crossover transition 
in $\xi$. It takes place  at the boundary of the weak and strong coupling domains. 
At strong coupling  occurring as one  reduces  $\sqrt{m\mu}$,
a dual description applies, in terms of a weakly coupled  non-Abelian  infrared free SQCD with 
the dual gauge group  U$(\tN)\times$U(1)$^{N-\tN}$ and  $N_f$ light dyon flavors.
The dual gauge group is Higgsed too (with a global color-flavor locked symmetry
preserved) and  supports non-Abelian strings. These strings  still confine {\em monopoles}
rather than quarks. 

Next, we increase the deformation parameter $\mu$  decoupling the
adjoint fields and sending the original theory  to the limit of \none SQCD. 
At large $\mu$ the dual theory is demonstrated to be   weakly coupled and infrared  free, with the 
 U$(\tN)$  gauge group  and $N_f$ light dyons $D^{lA}$, ($l=1,...\tN$ is the color index in the dual gauge group, while $A=1,...,N_f$ is the flavor index). Our 
proof is valid  provided the dyon condensate 
$\sim \xi \sim \sqrt{\mu m}$ is small enough.  
Non-Abelian strings (albeit this time non-BPS saturated)
are formed which confine monopoles --- qualitatively the same type of  confinement as in the \ntwo duality 
\cite{SYdual,SYtorkink}.
In the domain of small 
$\sqrt{\mu m}$ quark and gauge bosons of  original \none SQCD  are presented by stringy mesons built from the
monopole-antimonopoles pairs connected by two non-Abelian strings, see Fig.~\ref{figmeson}.

An interesting aspect, to be discussed in the bulk of the paper,
 is the relationship of our duality with that of Seiberg.
To make ourselves clear in this point we should undertake a small digression in the issue of vacua.

\none SQCD with $N_f$ flavors ($N+1\leq N_f<\frac{3}{2}N$) has a large number of 
distinct vacua. We need to  classify them. 
To this end we can invoke our knowledge of the vacuum structure in related theories, such as \ntwo SQCD, which is controlled by the exact Seiberg--Witten solution \cite{SW1,SW2}.

Let us turn to the latter. Among others, it has $N$ supersymmetric vacua which are generically referred to as the
``monopole vacua." The gauge symmetry is spontaneously broken down to ${\rm U}(1)^{N-1}$ in these 
vacua,\footnote{In our theory we will have ${\rm U}(1)^{N}$, since instead of the Seiberg--Witten SU$(N)$
group we work with  U$(N)$.}
and the subsequent switch-on of a small-$\mu$ deformation leads to the monopole condensation (in fact, in some of these vacua it is dyons that condense), the (Abelian) flux tube formation and confinement of (probe) quarks. As $\mu$ grows and eventually becomes large, the adjoint fields of the \ntwo theory decouple, and we are left with \none SQCD. 
The $N$ monopole vacua go though a crossover transition into a non-Abelian phase. We will say that
the above  vacua evolve and become the monopole vacua of \none SQCD. The name  ``monopole" is symbolic. 
We just continue to refer in this way to the vacua which used to be the monopole vacua of the \ntwo 
Seiberg--Witten
theory at small $\mu$, into the domain of large $\mu$ where the Seiberg--Witten control over dynamics is lost. 

At large $\mu$ we recover \none SQCD. The $N$ monopole vacua are those in which Seiberg's duality was established \cite{Sdual,IS}. If the quark fields of the electric theory are endowed with small masses to lift the continuous vacuum manifold, in the Seiberg magnetic dual theories the meson field $M$ condenses. Since it is singlet with respect to the dual color gauge groups SU($\tN$), this gauge group remains unbroken. We stress that the 
Seiberg $M$ condensation occurs in the vacua of the dual theory if in the original electric theory we stay in the monopole vacua. There is no obvious connection between $M$ and the monopole fields which --- the monopole fields --- are not defined at all  in this set up.\footnote{
A side remark which will not be elaborated below is in order here.
In \none SQCD with nonvanishing quark masses the Intriligator--Seiberg--Shih (ISS) vacuum was detected in 2006 
\cite{ISS,IS2}. With a generic set of the mass terms the ISS vacuum is non-supersymmetric (i.e. its energy is lifted from zero).
Given a special choice of the mass parameters it can be made supersymmetric at the {\em classical} level. 
Then, quantum corrections will lift it from zero, albeit the breaking can be small. In \cite{Shifman:2007kd} 
we considered non-Abelian strings and their junctions  in the ISS-like vacuum of \none SQCD.
Finally, once we started speaking of Seiberg's duality beyond the
original Seiberg's duality, we cannot help mentioning an inspiring paper of Komargodski \cite{komar}.}

Our task is to explore dualities in \none SQCD in the vacua other than the $N$ monopole vacua. 
For a deeper understanding of the problem we start, however,  
from the {\em quark vacua} of the \ntwo Seiberg--Witten theory 
(with addition of the the U(1) gauge group and the  corresponding 
Fayet--Iliopoulos term  \cite{FI}). This was the beginning of our program of
duality explorations,  the \ntwo limit \cite{SYdual,SYtorkink,SYcross,SYcrossp,SYrev}.
In this paper we report the study of the 
$\mu$-deformation leading us away from \ntwo to  \none SQCD, 
remaining in the quark vacua. 

Now, turning to 
 the relation between our duality (plus monopole confinement) and that of Seiberg  \cite{Sdual,IS},
 we observe that the  light dyons $D^{lA}$ of our U($\tN$) dual theory  are simultaneously
 similar to and dissimilar from Seiberg's ``dual quarks.'' 
 They have the same quantum numbers, but dynamics are different. One can conjecture that, 
 in fact, Seiberg's dual quarks
 is a different-phase implementation of the dyons $D^{lA}$. If so everything else becomes clear. 
Indeed, in quantum numbers, the stringy mesons formed from
the monopole-antimonopole pairs correspond to Seiberg's neutral mesons $M_A^B$ ($A,B=1,...,N_f$).
Both incorporate the singlet and adjoint representations of the global flavor group.
The difference is that in our dual theory these stringy mesons are nonperturbative objects which are 
rather heavy in the weak coupling
regime of the dual theory. This is in sharp contrast with the fact that, in Seiberg's dual theory, the $M_A^B$ mesons
appear as fundamental fields at the Lagrangian level and are light. 

The explanation for this dynamical differences was, in fact, given 
above: Seiberg's duality refers to the monopole vacua while ours to the quark vacua (of the $r=N$ type).
 Dyons $D^{lA}$ do not condense in the monopole vacua, and the (infrared
free) dual theory is in the  Coulomb phase.
In our vacua, at  
strong coupling  (weak coupling in the dual theory),
the light dyons condense, triggering   formation of the
non-Abelian strings and, as a result, the  confinement of monopoles. The dyon condensate
is proportional to $\sqrt{\mu m}$ and represents a would-be run-away vacuum not seen in the Seiberg
dual description, where $\mu$ is considered to be strictly  infinite (see Fig. \ref{figmuevol} in Section \ref{Seiberg}).

Concluding the introductory section we reiterate that the overall picture of duality we obtained
in previous works \cite{SYdual,SYtorkink} in \ntwo theories survives the passage to \none. Some 
details change, for instance, the strings cease
to be BPS-saturated (correspondingly, supersymmetry on the string world sheet is lost at the classical level). Nevertheless, the general pattern of the phenomenon stays {\em intact}.

The paper is organized as follows. In Sec. \ref{bulk} we outline our basic setup, \ntwo SQCD. Then we introduce a  small
deformation parameter $\mu$ and briefly review non-Abelian duality observed in \cite{SYdual,SYtorkink}. In 
Sec. \ref{N2duality} we describe
how the quarks and gauge bosons of the original theory pass into the monopole-antimonopole 
pairs, stringy mesons, 
in the crossover domain. In Sec. \ref{largemu} we increase $\mu$ eventually decoupling gauge singlets 
and adjoint scalars of the dual theory. This is the limit of \none SQCD.
Section \ref{largemustr} is devoted to formation of non-Abelian strings and monopole-antimonopole 
mesons in the \none theory. Then we proceed to duality in the quark vacua of the \none theory.
In Sec. \ref{Seiberg}  we compare  our picture to that of Seiberg. Finally, Sec. 7 summarizes our results and
conclusions. In Appendix we treat technical details of the U(3) model with $N_f=5$ at small $\mu$.

\section {Basic theory at small $\mu$}
\label{bulk}
\setcounter{equation}{0}

This section presents our basic  setup at small $\mu$, i.e  near the \ntwo limit.

The gauge symmetry of the basic bulk model is 
U($N$)=SU$(N)\times$U(1). In the absence
of  deformation the model under consideration is \ntwo  SQCD
 with $N_f$ massive quark hypermultiplets. 
 We assume that
$N_f>N$ but $N_f<\frac32 N$, see Eq.~(\ref{irfr}). 
The latter inequality ensures  that the dual theory is  {\em not} asymptotically free. 

In addition, we will introduce the mass term $\mu$ 
for the adjoint matter breaking \ntwo supersymmetry down to \none. 

The field content is as follows. The \ntwo vector multiplet
consists of the  U(1)
gauge field $A_{\mu}$ and the SU$(N)$  gauge field $A^a_{\mu}$,
where $a=1,..., N^2-1$, and their Weyl fermion superpartners plus
complex scalar fields $a$, and $a^a$ and their Weyl superpartners, respectively.
The $N_f$ quark multiplets of  the U$(N)$ theory consist
of   the complex scalar fields
$q^{kA}$ and $\tilde{q}_{Ak}$ (squarks) and
their   fermion superpartners --- all in the fundamental representation of 
the SU$(N)$ gauge group.
Here $k=1,..., N$ is the color index
while $A$ is the flavor index, $A=1,..., N_f$. We will treat $q^{kA}$ and $\tilde{q}_{Ak}$
as rectangular matrices with $N$ rows and $N_f$ columns. 

Let us first discuss the undeformed  \ntwo theory.
 The  superpotential has the form
 \beq
{\mathcal W}_{{\mathcal N}=2} = \sqrt{2}\,\sum_{A=1}^{N_f}
\left( \frac{1}{ 2}\,\tilde q_A {\mathcal A}
q^A +  \tilde q_A {\mathcal A}^a\,T^a  q^A\right)\,,
\label{superpot}
\eeq
where ${\mathcal A}$ and ${\mathcal A}^a$ are  chiral superfields, the ${\mathcal N}=2$
superpartners of the gauge bosons of  U(1) and SU($N$), respectively.

Next, we add the mass term for the adjoint fields which breaks \ntwo
supersymmetry down to \none,
\beq
{\mathcal W}_{[\mu ]}=\sqrt{\frac{N}{2}}\,\frac{\mu_1}{2} {\mathcal A}^2
+  \frac{\mu_2}{2}({\mathcal A}^a)^2,
\label{msuperpotbr}
\eeq
where $\mu_1$ and $\mu_2$ is are mass parameters for the chiral
superfields in \ntwo gauge supermultiplets,
U(1) and SU($N$), respectively. Generally speaking $\mu_1$ need not coincide with $\mu_2$,
but we will assume these parameters to be of the same order of magnitude and will generically denote them as $\mu$.
Clearly, the mass term (\ref{msuperpotbr}) splits the 
\ntwo supermultiplets, breaking
\ntwo supersymmetry down to \none. First we assume that $\mu$ is small, much smaller than
the quark masses $m_A$,
\beq
\mu\ll m_A, \qquad A=1,...,N_f\,.
\label{smallmu}
\eeq
The bosonic part of the action of our basic 
theory has the form  (for details see \cite{SYrev})
\beqn
S&=&\int d^4x \left[\frac1{4g^2_2}
\left(F^{a}_{\mu\nu}\right)^2 +
\frac1{4g^2_1}\left(F_{\mu\nu}\right)^2
+
\frac1{g^2_2}\left|D_{\mu}a^a\right|^2 +\frac1{g^2_1}
\left|\partial_{\mu}a\right|^2 \right.
\nonumber\\[4mm]
&+&\left. \left|\nabla_{\mu}
q^{A}\right|^2 + \left|\nabla_{\mu} \bar{\tilde{q}}^{A}\right|^2
+V(q^A,\tilde{q}_A,a^a,a)\right]\,.
\label{model}
\eeqn
Here $D_{\mu}$ is the covariant derivative in the adjoint representation
of  SU$(N)$, while
\beq
\nabla_\mu=\partial_\mu -\frac{i}{2}\; A_{\mu}
-i A^{a}_{\mu}\, T^a
\eeq
\label{defnabla}
acts in the fundamental representation.
We suppress the color  SU($N$)  indices of the matter fields. The normalization of the 
 SU($N$) generators  $T^a$ is as follows
$$
{\rm Tr}\, (T^a T^b)=\mbox{$\frac{1}{2}$}\, \delta^{ab}\,.
$$
The coupling constants $g_1$ and $g_2$
correspond to the U(1)  and  SU$(N)$  sectors, respectively.
With our conventions, the U(1) charges of the fundamental matter fields
are $\pm1/2$, see Eq.~(\ref{defnabla}).

The scalar potential $V(q^A,\tilde{q}_A,a^a,a)$ in the action (\ref{model})
is the  sum of  $D$ and  $F$  terms,
\beqn
V(q^A,\tilde{q}_A,a^a,a) &=&
 \frac{g^2_2}{2}
\left( \frac{1}{g^2_2}\,  f^{abc} \bar a^b a^c
 +
 \bar{q}_A\,T^a q^A -
\tilde{q}_A T^a\,\bar{\tilde{q}}^A\right)^2
\nonumber\\[3mm]
&+& \frac{g^2_1}{8}
\left(\bar{q}_A q^A - \tilde{q}_A \bar{\tilde{q}}^A \right)^2
\nonumber\\[3mm]
&+& 2g^2_2\left| \tilde{q}_A T^a q^A 
+\frac{1}{\sqrt{2}}\,\,\frac{\pt{\mathcal W}_{\mu}}{\pt a^a}\right|^2+
\frac{g^2_1}{2}\left| \tilde{q}_A q^A +\sqrt{2}\,\,\frac{\pt{\mathcal W}_{\mu}}{\pt a} \right|^2
\nonumber\\[3mm]
&+&\frac12\sum_{A=1}^{N_f} \left\{ \left|(a+\sqrt{2}m_A +2T^a a^a)q^A
\right|^2\right.
\nonumber\\[3mm]
&+&\left.
\left|(a+\sqrt{2}m_A +2T^a a^a)\bar{\tilde{q}}^A
\right|^2 \right\}\,.
\label{pot}
\eeqn
Here  $f^{abc}$ denote the structure constants of the SU$(N)$ group,
$m_A$ is the mass term for the $A$-th flavor,
 and 
the sum over the repeated flavor indices $A$ is implied.

\subsection{Vacuum structure}

Now, let us discuss  the vacuum structure of  this theory \cite{SYfstr}.
The  vacua of the theory (\ref{model}) are determined by the zeros of 
the potential (\ref{pot}). In general, the theory has a number of the so called $r$-vacua, in which 
 $r$ quarks condense. The range of variation of $r$ is  $r=0,...,N$. Say,
$r=0$ vacua (there are $N$ such vacua)  are always at strong
coupling. We have already explained that they are called the 
monopole vacua \cite{SW1,SW2}. In this paper we will focus
on a particular set of vacua with the maximal number of condensed quarks, $r=N$.
The reason for this choice is that all U(1) factors of the gauge group are spontaneously 
broken in these vacua, and, as a result, 
they support non-Abelian strings \cite{HT1,ABEKY,SYmon,HT2}. The occurrence of strings ensures 
the monopole confinement in these vacua.

Let us first assume  that our theory is at weak coupling,  so that we can  
analyze it quasiclassically. Below we will explicitly formulate  conditions on the quark mass terms
and $\mu$ which will enforce such a regime.

With generic values of the quark masses we have 
\beq
C_{N_f}^{N}= \frac{N_f!}{N!(N_f-N)!}
\label{numva}
\eeq
 isolated $r$-vacua in which $r=N$ quarks (out of $N_f$) develop
vacuum expectation values  (VEVs).
Consider, say, the vacuum in which the first $N$ flavors develop VEVs, to be denoted as (1, 2 ..., $N$).
In this vacuum  the
adjoint fields  develop  
VEVs too, namely,
\beq
\left\langle \left(\frac12\, a + T^a\, a^a\right)\right\rangle = - \frac1{\sqrt{2}}
 \left(
\begin{array}{ccc}
m_1 & \ldots & 0 \\
\ldots & \ldots & \ldots\\
0 & \ldots & m_N\\
\end{array}
\right),
\label{avev}
\eeq
For generic values of the quark masses, the  SU$(N)$ subgroup of the gauge 
group is
broken down to U(1)$^{N-1}$. However, in the {\em special limit}
\beq
m_1=m_2=...=m_{N_f},
\label{equalmasses}
\eeq
the  adjoint field VEVs do not break the SU$(N)\times$U(1) gauge group.
In this limit the theory acquires  a global flavor SU$(N_f)$ symmetry.

With all quark masses equal and to the leading order in $\mu$,
 the mass term for the adjoint matter (\ref{msuperpotbr})
reduces to the Fayet--Iliopoulos $F$-term of the U(1) factor of the SU$(N)\times$U(1) gauge group,
  which does {\em not} break \ntwo supersymmetry \cite{HSZ,VY}. In this limit the Fayet--Iliopoulos $F$-term
can be transformed into the Fayet--Iliopoulos $D$-term by an SU$(2)_R$ rotation; the theory reduces to
\ntwo SQCD described in detail, say, in \cite{SYrev}.
Higher orders in the parameter $\mu$
  break \ntwo supersymmetry by splitting all \ntwo multiplets.

If the quark masses are unequal the U($N$) gauge group is broken  down to U(1)$^{N}$
by the adjoint field VEV's (\ref{avev}). To the leading order in $\mu$, the superpotential
(\ref{msuperpotbr}) reduces to $N$ distinct FI terms, one in each U(1) gauge factor.
\ntwo supersymmetry in each individual low-energy U(1) theory remains unbroken \cite{SYfstr}. It is
broken, however, being considered in the full U($N$) gauge theory.

Using (\ref{msuperpotbr}) and (\ref{avev}) it is not difficult to obtain the quark field VEVs
 from Eq.~(\ref{pot}).   By virtue of a gauge rotation they can be written as
\beqn
\langle q^{kA}\rangle &=& \langle\bar{\tilde{q}}^{kA}\rangle=\frac1{\sqrt{2}}\,
\left(
\begin{array}{cccccc}
\sqrt{\xi_1} & \ldots & 0 & 0 & \ldots & 0\\
\ldots & \ldots & \ldots  & \ldots & \ldots & \ldots\\
0 & \ldots & \sqrt{\xi_N} & 0 & \ldots & 0\\
\end{array}
\right),
\nonumber\\[4mm]
k&=&1,..., N\,,\qquad A=1,...,N_f\, ,
\label{qvev}
\eeqn
where we present the quark fields as  matrices in the color ($k$) and flavor ($A$) indices.
The Fayet--Iliopoulos $F$-term parameters for each U(1) gauge factor are given (in the quasiclassical
approximation) by the following expressions:
\beq
\xi_P = 2\left\{\sqrt{\frac{2}{N}}\,\,\,\mu_1\,\hat{m}+\mu_2(m_P-\hat{m})\right\},
\qquad P=1,...,N
\label{xis}
\eeq
and $\hat{m}$ is the average value of the first $N$ quark masses,
\beq
\hat{m}=\frac1{N}\sum_{P=1}^{N} m_P\, .
\label{avm}
\eeq

While the adjoint VEVs do not break the SU$(N)\times$U(1) gauge group in the limit
(\ref{equalmasses}), the quark condensate (\ref{qvev}) does result in  the spontaneous
breaking of both gauge and flavor symmetries.
A diagonal global SU$(N)$ combining the gauge SU$(N)$ and an
SU$(N)$ subgroup of the flavor SU$(N_f)$
group survives, however. This is color-flavor locking. Below we will refer to this diagonal
global symmetry as to $ {\rm SU}(N)_{C+F}$.

Thus, the pattern  of the
color and flavor symmetry breaking
is as follows: 
\beq
{\rm U}(N)_{\rm gauge}\times {\rm SU}(N_f)_{\rm flavor}\to  
{\rm SU}(N)_{C+F}\times  {\rm SU}(\tilde{N})_F\times {\rm U}(1)\,,
\label{c+f}
\eeq
where $\tilde{N}=N_f-N$.
Here SU$(N)_{C+F}$ is a global unbroken color-flavor rotation, which involves the
first $N$ flavors, while the SU$(\tN )_F$ factor stands for the flavor rotation of the 
$\tN$ quarks.
The presence of the global SU$(N)_{C+F}$ group is instrumental for
formation of the non-Abelian strings \cite{HT1,ABEKY,SYmon,HT2,SYfstr}.
As we will see shortly, the global symmetry of the dual theory is, of course, 
the same, albeit the physical origin is different.

With unequal quark masses, the  global symmetry  (\ref{c+f}) is broken down to 
U(1)$^{N_f-1}$ both by the adjoint and squark VEVs. This should be contrasted with the 
theory with the Fayet--Iliopoulos term introduced through the $D$-term,  in which the
quark VEVs are all equal and do not break the color-flavor symmetry.

Since the global (flavor) SU$(N_f)$ group is broken by the quark VEVs anyway, it will be helpful for 
our purposes
to consider
the following mass splitting:
\beq
m_P=m_{P'}, \qquad m_K=m_{K'}, \qquad m_P-m_K=\Delta m
\label{masssplit}
\eeq
where 
\beq
 P, P'=1, ... , N\,\,\,\, {\rm and}   \,\,\,\, K, K'=N+1, ... , N_f\,.
 \label{pppp}
\eeq
This mass splitting respects the global
group (\ref{c+f}) in the $(1,2,...,N)$ vacuum. Moreover, this vacuum  becomes  isolated.
No Higgs branch  develops.  We will often use this limit below.

\subsection{Perturbative excitations}

Now let us discuss the  perturbative excitation spectrum. 
To the leading order in $\mu$,  in the limit (\ref{masssplit}), the superpotential
(\ref{msuperpotbr})  reduces to the Fayet--Iliopoulos $F$-term of the U(1) factor of the  gauge group. Since
both U(1) and SU($N$) gauge groups are broken by the squark condensation, all
gauge bosons become massive. In fact, with nonvanishing $\xi_P$'s (see Eq.~(\ref{xis})), both the quarks and adjoint scalars  
combine  with the gauge bosons to form long \ntwo supermultiplets \cite{VY},  for a review see \cite{SYrev}.
In the limit (\ref{masssplit}) $$\xi_P\equiv\xi\,,$$  and all states come in 
representations of the unbroken global
 group (\ref{c+f}), namely, in the singlet and adjoint representations
of SU$(N)_{C+F}$,
\beq
(1,\, 1), \quad (N^2-1,\, 1),
\label{onep}
\eeq
 and in the bifundamental representations
\beq
 \quad (\bar{N},\, \tN), \quad
(N,\, \bar{\tN})\,.
\label{twop}
\eeq
We mark representations in (\ref{onep}) and (\ref{twop})  with respect to two 
non-Abelian factors in (\ref{c+f}). The singlet and adjoint fields are (i) the gauge bosons, and
(ii) the first $N$ flavors of the squarks $q^{kP}$ ($P=1,...,N$), together with their fermion superpartners.
The bifundamental fields are the quarks $q^{kK}$ with $K=N+1,...,N_f$.
These quarks transform in the two-index representations of the global
group (\ref{c+f}) due to the color-flavor locking.

In the limit (\ref{masssplit}) the mass of the $(N^2-1,\, 1)$ adjoint fields is 
\beq
m_{(N^2-1,\,1)}=g_2\sqrt{\xi}\,,
\label{Wmass}
\eeq
while the singlet field  mass is
 \beq
m_{(1,\,1)}=g_1\, \sqrt{\frac{N}{2}}\,\sqrt{\xi}\,.
\label{phmass}
\eeq
The bifundamental field masses are given by
\beq
m_{(\bar{N},\, \tN)}= \Delta m\,.
\label{bifund}
\eeq 

The above quasiclassical analysis is valid if the theory is at weak coupling. This is the case if
the quark VEVs are sufficiently large so that the  gauge coupling constant is frozen at a large scale.
From (\ref{qvev}) we see that 
the quark condensates are of the order of
$\sqrt{\mu m}$ (see also \cite{SW1,SW2,APS,CKM}). As was mentioned, we assume that $\mu_1\sim\mu_2\sim\mu$. In this case the weak 
coupling condition reduces to
\beq
\sqrt{\mu m}\gg\Lambda_{{\mathcal N}=2}\,,
\label{weakcoup}
\eeq
where $\Lambda_{{\mathcal N}=2}$ is the scale of the \ntwo  theory, and we assume that all quark masses are of the same order $m_A\sim m$. 
In particular,  the condition (\ref{weakcoup}), combined with the condition (\ref{smallmu}) of smallness
of $\mu$,   implies that the average quark mass $m$ is very large.

\section{Duality at small $\mu$ in the quark vacua }
\label{N2duality}
\setcounter{equation}{0}

\subsection{Dual theory}

Now we will relax the condition (\ref{weakcoup}) and pass
to the strong coupling domain at 
\beq
|\sqrt{\xi_P}|\ll \Lambda_{{\mathcal N}=2}\,, \qquad | m_{A}|\ll \Lambda_{{\mathcal N}=2}\,.
\label{strcoup}
\eeq
 \ntwo SQCD with the Fayet--Iliopoulos term (introduced as the $D$-term) was shown \cite{SYdual,SYtorkink}  
 to undergo a
crossover transition upon reduction  of the
FI parameter.   The results obtained in   \cite{SYdual} are based on
studying the Seiberg--Witten curve \cite{SW1,SW2,APS} in \ntwo SQCD on the Coulomb branch, and,
therefore, do not depend on the type of the FI deformation. We briefly review these results here
adjusting our consideration  \cite{SYdual,SYtorkink} to fit the case  of the Fayet--Iliopoulos  $F$-term 
induced by the adjoint mass  $\mu$. 

The  domain (\ref{strcoup}) 
can be described in terms of weakly coupled (infrared free) dual theory
with with the gauge group
\beq
{\rm U}(\tN)\times {\rm U}(1)^{N-\tN}\,,
\label{dualgaugegroup}
\eeq
 and $N_f$ flavors of light {\em dyons}.\footnote{ Previously the SU$(\tN)$
 gauge group  was identified as dual  \cite{APS} on the Coulomb   branch 
at the root of the baryonic Higgs branch in the \ntwo supersymmetric  SU($N$) Yang--Mills 
theory with massless quarks.}

\vspace{2mm}

Light dyons $D^{lA}$ 
\beq
l=1, ... ,\tN\,, \quad A=1, ... , N_f
\label{ldnumb}
\eeq 
are in 
the fundamental representation of the gauge group
SU$(\tN)$ and are charged under the Abelian factors indicated in Eq.~(\ref{dualgaugegroup}).
 In addition, there are  $(N-\tN)$ 
light dyons $D^J$ ($J=\tN+1, ..., N$), neutral under 
the SU$(\tN)$ group, but charged under the
U(1) factors. 

In Appendix A we present the low-energy effective action for 
the dual theory in a specific 
example:  $N=3$, $N_f=5$,  and $\tN=2$. In particular, starting from this action, we find
the dyon condensates in the quasiclassical approximation.
Generalization of these results to arbitrary $N$ and $\tN$ has the following form
\beqn
\langle D^{lA}\rangle \! \! &=& \langle \bar{\tilde{D}}^{lA}\rangle =
\!\!
\frac1{\sqrt{2}}\,\left(
\begin{array}{cccccc}
0 & \ldots & 0 & \sqrt{\xi_{1}} & \ldots & 0\\
\ldots & \ldots & \ldots  & \ldots & \ldots & \ldots\\
0 & \ldots & 0 & 0 & \ldots & \sqrt{\xi_{\tN}}\\
\end{array}
\right),
\nonumber\\[4mm]
\langle D^{J}\rangle &=& \langle\bar{\tilde{D}}^{J}\rangle=\sqrt{\frac{\xi_J}{2}}, 
\qquad J=\tN +1, ..., N\,.
\label{Dvev}
\eeqn

The most important feature apparent in (\ref{Dvev}), as compared to the squark VEVs  of the 
original theory (\ref{qvev}),  is a ``vacuum jump'' \cite{SYdual},
\beq
(1, ... ,\, N)_{\sqrt{\xi}\gg \Lambda_{{\mathcal N}=2}} \to (N+1, ... , \,N_f,\,\,\tN+1, ... ,\, N)_{\sqrt{\xi}\ll \Lambda_{{\mathcal N}=2}}\,.
\label{jump}
\eeq
In other words, if we pick up the vacuum with nonvanishing VEVs of the  first $N$ quark flavors
in the original theory at large $\xi$, Eq.~(\ref{model}),  and then reduce $\xi$ below 
$\Lambda_{{\mathcal N}=2}$, 
the system goes through a crossover transition and ends up in the vacuum of the {\em dual} theory with
the nonvanishing VEVs of $\tN$ last dyons (plus VEVs of $(N-\tN)$ SU$ (\tN)$ singlets).

The Fayet--Iliopoulos parameters $\xi_P$  in (\ref{Dvev}) are determined by the 
quantum version of the classical expressions
(\ref{xis}).  They are expressible  via the roots of the Seiberg--Witten curve 
in the given $r=N$ vacuum \cite{SYfstr}. 
Namely,
\beq
\xi_P=2\,\left\{\sqrt{\frac{2}{N}}\,\mu_1\,\hat{m}-\mu_2(\sqrt{2}e_P+\hat{m})\right\},
\label{qxis}
\eeq
where $e_P$ are the double roots of the Seiberg--Witten curve \cite{APS}, 
\beq
y^2= \prod_{P=1}^{N} (x-\phi_P)^2 -
4\left(\frac{\Lambda}{\sqrt{2}}\right)^{N-\tN}\, \,\,\prod_{A=1}^{N_f} \left(x+\frac{m_A}{\sqrt{2}}\right),
\label{curve}
\eeq
while $\phi_P$ are gauge invariant parameters on the Coulomb branch. We recall that $\hat{m}$
in Eq.~(\ref{qxis}) is still the average of the first $N$ quark masses (\ref{avm}).
The   curve (\ref{curve}) describes the Coulomb branch of the theory for  $\tN<N-1$. The case  $   \tN=N-1$ is special. In this case we must make a shift in Eq.~(\ref{curve}) \cite{APS},
\beq
m_A\to \tilde{m}_A=m_A+\frac{\Lambda_{{\mathcal N}=2}}{N}, \qquad \tN=N-1\,.
\label{shift}
\eeq
We will not consider this special case at large $\mu$ since it is incompatible with the 
condition $N_f<3/2\, N$ or $\tN< N/2$.

In the $r=N$ vacuum the curve (\ref{curve}) has $N$ double roots and reduces to
\beq
y^2= \prod_{P=1}^{N} (x-e_P)^2,
\label{rNcurve}
\eeq
where quasiclassically (at large masses) $e_P$'s are given  by the
mass parameters, $\sqrt{2}e_P\approx -m_P$, $P=1,...,N$.

As long as we keep $\xi_P$ and masses small enough (i.e. in the domain (\ref{strcoup}))
the coupling constants of the
infrared free dual theory (frozen at the scale of the dyon VEVs) are small:
the dual theory is at weak coupling.

At small masses, in the region  (\ref{strcoup}), the double roots of the Seiberg--Witten
 curve are  
\beq
\sqrt{2}e_I = -m_{I+N}, \qquad 
\sqrt{2}e_J = \Lambda_{{\mathcal N}=2}\,\exp{\left(\frac{2\pi i}{N-\tN}J\right)}
\label{roots}
\eeq
for $\tN<N-1$, where 
\beq
I=1, ... ,\tN\,\,\,\, {\rm  and} \,\,\,\, J=\tN+1, ... , N\,.
\label{d1}
\eeq
 In particular, the $\tN$ first roots are determined by the masses of the 
 last $\tN$ quarks --- a reflection of the fact that the 
non-Abelian sector of the dual theory is not asymptotically free and is at weak coupling
in the domain (\ref{strcoup}). 

From Eqs.~(\ref{Dvev}), (\ref{qxis}) and (\ref{roots}) we see that the VEVs of the non-Abelian
dyons $D^{lA}$ are determined by $\sqrt{\mu m}$ and are much smaller 
than the VEVs of the Abelian dyons $D^{J}$ in the domain
(\ref{strcoup}).  The latter are of the order of 
$\sqrt{\mu \Lambda_{{\mathcal N}=2}}$. 
This circumstance is most crucial for our analysis in this paper. It will allow us to eventually increase $\mu$
and decouple the adjoint fields without spoiling the weak coupling condition in the dual theory,
see Sec.~\ref{largemu}.

In the special case $\tN=N-1$ masses in (\ref{roots}) should be shifted according to
(\ref{shift}).

Now, let us consider either equal quark masses or the mass choice (\ref{masssplit}).
Both, the gauge group and the global flavor SU($N_f$) group, are
broken in the vacuum. However, the color-flavor locked form inherent to (\ref{Dvev})
under the given mass choice guarantees that the diagonal
global SU($\tN)_{C+F}$ symmetry survives. More exactly, the  unbroken {\em global} group of the dual
theory is 
\beq
 {\rm SU}(N)_F\times  {\rm SU}(\tN)_{C+F}\times {\rm U}(1)\,.
\label{c+fd}
\eeq
The SU$(\tN)_{C+F}$ factor in (\ref{c+fd}) is a global unbroken color-flavor rotation, which involves the
last $\tN$ flavors, while the SU$(N)_F$ factor stands for the flavor rotation of the 
first $N$ dyons. 

Thus, a color-flavor locking takes place in the dual theory too. Much in the same way as 
in the original theory, the presence of the global SU$(\tN)_{C+F}$ group 
is the  reason behind formation of the non-Abelian strings.
 For generic quark masses the  global symmetry  (\ref{c+f}) is broken down to 
U(1)$^{N_f-1}$. 

In the equal mass limit, or given the mass choice (\ref{masssplit}),
the global unbroken symmetry (\ref{c+fd}) of the dual theory at small
$\xi$ coincides with the global group (\ref{c+f}) which manifest in the
$r=N$ vacuum of the original theory at large
$\xi$.  This has been already announced previously.

Note, however, that this global symmetry is realized in two very distinct ways in the dual pair at hand.
As was already mentioned, the quarks and U($N$) gauge bosons of the original theory at large $\xi$
come in the $(1,1)$, $(N^2-1,1)$, $(\bar{N},\tN)$, and $(N,\bar{\tN})$
representations of the global group (\ref{c+f}), while the dyons and U($\tN$) gauge 
bosons form $(1,1)$, $(1,\tN^2-1)$, $(N,\bar{\tN})$, and 
$(\bar{N},\tN)$ representations of (\ref{c+fd}). We see that the
adjoint representations of the $(C+F)$
subgroup are different in two theories. How can this happen?
     
\begin{figure}
\epsfxsize=7cm
\centerline{\epsfbox{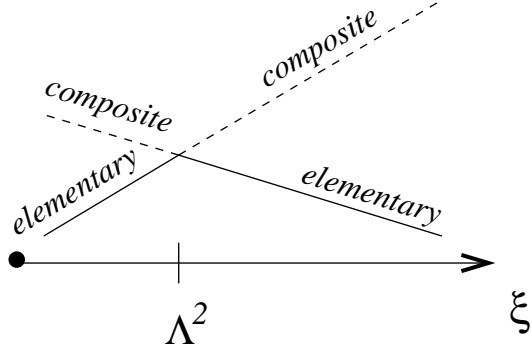}}
\caption{\small Evolution of the SU$(N)$ and SU$(\tN)$ gauge bosons and light quarks (dyons) vs. $\xi$. 
}
\label{figevol}
\end{figure}

The quarks and gauge bosons
which form the  adjoint $(N^2-1)$ representation  
of SU($N$) at large $\xi$ and the dyons and gauge bosons which form the  adjoint $(\tN^2-1)$ representation  of SU($\tN$) at small $\xi$ are, in fact, {\em distinct} states.
The $(N^2-1)$  adjoints of SU($N$) become heavy 
and decouple as we pass from large to small $\xi$ 
 along the line $\xi\sim \Lambda_{{\mathcal N}=2}$. Moreover, some 
composite $(\tN^2-1)$ adjoints  of SU($\tN$), which are 
heavy  and invisible in the low-energy description at large $\xi$ become light 
at small $\xi$ and form the $D^{lK}$ dyons
 ($K=N+1,...,N_f$) and gauge bosons of U$(\tN)$. The phenomenon of 
 the level crossing
 takes place (Fig.~\ref{figevol}). Although this crossover is smooth in the full theory,
from the standpoint of the low-energy description the passage from  large to small $\xi$  means a dramatic change: the low-energy theories in these domains are 
completely
different; in particular, the degrees of freedom in these theories are different.

This logic leads us to the following conclusion. In addition to light dyons and gauge bosons 
included in  the low-energy theory at small $\xi$ we must have
heavy  fields  which form the adjoint representation $(N^2-1,1)$ of the 
global symmetry (\ref{c+fd}). These are screened  quarks 
and gauge bosons from the large-$\xi$ domain.
Let us denote them as $M_P^{P'}$ (here $P , P'=1, ... , N$). 

As has been already noted in Sec.~\ref{intro},
at small $\xi$ they decay into the monopole-antimonopole 
pairs on the curves of marginal stability (CMS).\footnote{An explanatory remark regarding
our terminology is in order. Strictly speaking,
such pairs  can be  formed by monopole-antidyons and
 dyon-antidyons as well, the dyons carrying root-like electric charges. 
 In this paper we refer to all such states collectively as to
``monopoles." This is to avoid confusion with dyons which appear in Eq.~(\ref{Dvev}). The
latter dyons carry weight-like electric charges and, roughly speaking, behave as
quarks, see \cite{SYdual} for further details.}
This is in accordance with results obtained 
for \ntwo SU(2) gauge theories \cite{SW1,SW2,BF} on the Coulomb branch at zero $\xi$,
while for the theory at hand it is proven in \cite{SYtorkink}.
The general rule is that the only states which exist at strong coupling inside CMS are those which can become massless on the Coulomb branch
\cite{SW1,SW2,BF}. For our theory these are light dyons shown in Eq.~(\ref{Dvev}),
gauge bosons of the dual gauge group and monopoles.
 
 At small nonvanishing $\xi$ the
monopoles and antimonopoles produced in the decay process of the adjoint $(N^2-1,1)$ states
cannot escape from
each other and fly off to asymptotically large separations 
because they are confined. Therefore, the (screened) quarks or gauge bosons 
evolve into stringy mesons $M_P^{P'}$ ($P,P'=1, ..., N$) in the  strong coupling domain of small  $\xi$ 
-- the  monopole-antimonopole
pairs connected  by two strings \cite{SYdual,SYtorkink}, as shown in  Fig.~\ref{figmeson}. 

By the same token,  at large $\xi$, in addition to the light quarks and gauge bosons,
we  have heavy fields $M_K^{K'}$ (here $K, K'=N+1, ... ,  N_f$), which  form the  
adjoint $(\tN^2-1)$ representation  of SU($\tN$).
This is schematically depicted in Fig.~\ref{figevol}. 

The $M_K^{K'}$ states are (screened)  light
dyons and gauge bosons of the dual theory. In the large-$\xi$ domain they decay into
the monopole-antimonopole 
pairs and form stringy mesons  \cite{SYdual} shown in Fig.~\ref{figmeson}.

\subsection{More on the non-Abelian strings  and confined mono\-poles}
\label{stringsN2}

Since dyons  develop VEVs in the $r=N$ vacuum which break 
the gauge group, see (\ref{Dvev}), our dual theory  supports strings.
In fact, the minimal stings in our theory are the $Z_N$ strings, progenitors of 
the non-Abelian strings \cite{HT1,ABEKY,SYmon,HT2}.

At generic $m_A$ the dual gauge group (\ref{dualgaugegroup}) reduces to U(1)$^N$;
the low-energy theory is U(1)$^N$
gauge theory with the Fayet--Iliopoulos $F$-term for each U(1) factor.  The $Z_N$ strings for this 
theory are thoroughly studied in \cite{SYfstr}. In the low-energy approximation the $Z_N$ strings are BPS saturated.
Tensions of all $N$ elementary $Z_N$ strings are given by the FI parameters \cite{SYfstr},
\beq
T^{\rm BPS}_{P}=2\pi|\xi_P|, \qquad P=1,...,N\,.
\label{ten}
\eeq

In the limit (\ref{masssplit}) the color-flavor locking takes place and the 
global group of the dual theory becomes 
that of Eq.~(\ref{c+fd}). In this case $\tN$ of the set of $N$ $Z_N$ strings (associated with
windings of non-Abelian $D^{lA}$ dyons) acquire orientational 
zero modes and become non-Abelian. They can be analyzed  within the general framework developed in 
\cite{HT1,ABEKY,SYmon,HT2}, see \cite{SYrev} for a review. 
The internal dynamics of 
the orientational zero  modes are described by two-dimensional 
\ntwot supersymmetric CP model living on the string world sheet \cite{HT1,ABEKY,SYmon,HT2}. 
For the original theory (\ref{model}) it is CP$(N-1)$ model for $\tN =0$. For nonzero $\tN$ the string
becomes semilocal. Semilocal strings do not have fixed transverse radius, they acquire 
 size moduli, see    \cite{AchVas} for a  review of 
the Abelian semilocal strings. The non-Abelian semilocal strings in \ntwo SQCD with $N_f>N$ were studied in \cite{HT1,HT2,SYsem,Jsem}.
The internal dynamics of these strings is {\em qualitatively} described by a weighted
CP$(N_f-1)$ model with $N$ positive and $\tN$ negative charges associated with $N$ 
orientational moduli and $\tN$ size moduli. (Aspects of a quantitative description and 
its interrelation with the  weighted
CP$(N_f-1)$ model will be discussed in \cite{SYV}.)
In the dual  theory $N$ and $\tN$ interchange; it is governed by 
the weighted CP$(N_f-1)$ model with $\tN$ positive (orientations) and $N$ negative (size) charges
$n_K$, $K=(N+1),...,N_f$ and $\rho_P$, $P=1,...,N$, respectively \cite{SYtorkink}. 
The above moduli are subject to the constraint
\beq
  |n_K|^2 - |\rho_P|^2=2 \tilde{\beta},
\label{unitvec}
\eeq
where $\tilde{\beta}$ is a coupling constant of the dual world-sheet theory. It is determined by
the gauge coupling constant of the dual bulk theory at the scale $\sim \sqrt{\xi}$ 
\cite{ABEKY,SYmon,SYtorkink},
\beq
\tilde{\beta}= \frac{2\pi}{\tilde{g}_2^2}\,.
\label{betag}
\eeq

Distinct elementary non-Abelian strings 
correspond to different vacua of the CP model under consideration.
Confined  monopoles of the bulk theory are identified with the junctions of
two degenerate elementary non-Abelian strings \cite{T,SYmon,HT2}.
These are seen as kinks interpolating between different vacua of the CP model. Non-perturbative 
generation of the dynamical scale $\Lambda_{CP}$ in 
the CP model stabilizes these kinks in the non-Abelian regime,
 making their inverse sizes and masses of the order of $\Lambda_{CP}$ \cite{SYmon,SYrev}.
Thus, the notion of the confined monopole becomes well-defined in the non-Abelian limit.

The identification of confined monopoles with the  CP-model kinks  reveals the global
quantum numbers of the monopoles.  Say,  it was known for a long time that  
the kinks in the quantum limit form a fundamental representation
of the global SU$(N)$ group  in the  \ntwot supersymmetric \cpn models \cite{W79,HoVa}.
In \cite{SYtorkink} we generalized this result to the case of the 
\ntwot supersymmetric weighted CP models. We showed that 
the kinks (confined monopoles) are in the fundamental representation of the global group (\ref{c+fd}). 

More exactly, in the limit (\ref{masssplit}) they
form the $(N,1)+(1,\tN)$  representations of the global group (\ref{c+fd}).
This means that the total number of stringy mesons $M_A^B$ formed by 
the monopole-antimonopole
pairs connected by two different elementary non-Abelian strings  (Fig.~\ref{figmeson}) is  $N_f^2$.
The mesons $M_P^{P'}$ form the
singlet and $(N^2-1,1)$ adjoint representations  of the global group (\ref{c+fd}), 
the mesons $M_P^{K}$ and $M_K^{P}$ form bifundamental representations
$(N,\bar{\tN})$ and $(\bar{N},\tN)$, while the mesons 
$M_K^{K'}$ form the singlet and $(1,\tN^2-1)$ adjoint representations. 
(Here, as usual, $P=1,...,N$ and $K=(N+1) , ... , N_f$.)

All these mesons with not too high  spins  have masses 
\beq
m_{M_P^{P'}} \sim \sqrt{\xi}\, ,
\label{mstringyN2}
\eeq
as determined by the  string tensions (\ref{ten}) .
They are heavier than the
elementary states, namely, dyons and dual gauge bosons which form 
the (1,1), $(N,\bar{\tN})$, $(\bar{N},\tN)$, 
and $(1,\tN^2-1)$ representations and have masses $\sim \tilde{g}_2\sqrt{\xi}$.

Therefore, the  (1,1), $(N,\bar{\tN})$, $(\bar{N},\tN)$, and $(1,\tN^2-1)$ stringy mesons
decay into elementary states, and we are left with  $M_P^{P'}$ stringy mesons
in the representation $(N^2-1,1)$.
Thus our confinement picture  in the bulk theory outlined above is confirmed by the world-sheet analysis.

This concludes our extended introduction and adjustments necessary to pass to the study of 
the \none theories.

\section{Flowing to \none QCD}
\label{largemu}
\setcounter{equation}{0}

With all preparatory work done, we begin our journey in the \none theories.
In this section we increase  the adjoint mass   $\mu$ and decouple the adjoint matter. In the course of this process
the theory at hand flows to \none SQCD. So, now we assume that 
\beq
|\mu| \gg |m_A |, \qquad A=1, ... , N_f\,.
\label{mularge}
\eeq
Then, the \ntwo multiplets are split. We consider the
 quark masses to be small enough to guarantee that the original theory (\ref{model})
is at  strong coupling, while the dual theory is at weak coupling.

\subsection{Decoupling the U(1)$^{(N-\tN)}$ sector}
\label{41}

At first,  we will impose the condition
\beq
|\mu |\ll \Lambda_{{\mathcal N}=2}\,,
\label{intermu}
\eeq
implying (in conjunction with (\ref{mularge})) that all 
parameters $\sqrt{\xi_P}$ are much smaller than $\Lambda_{{\mathcal N}=2}$. 
Then our dual theory is at   weak coupling, see Eqs. (\ref{qxis}) and  (\ref{roots}). 
From (\ref{roots}) we see that 
VEVs of non-Abelian dyons $D^{lA}$ are much smaller those of the Abelian dyons $D^J$. Consider
the low-energy limit of the dual theory, i.e.  energies much lower than 
$\sqrt{\mu\Lambda_{{\mathcal N}=2}}$. In this scale the Abelian dyons $D^J$ 
$$J=(\tN+1), ... , N$$
are heavy and decouple. These dyons interact with $(N-\tN +1)$ U(1) gauge fields, see 
Eq.~(\ref{dualgaugegroup}). In this set of gauge bosons,  $(N-\tN)$ U(1) fields also become heavy (with masses 
$g\sqrt{\mu\Lambda_{{\mathcal N}=2}}$). Only one remains. As a result, in the low-energy limit we are
left with the dual theory with the gauge group
\beq
{\rm U}(\tN)
\label{dgg}
\eeq
 and $N_f$ flavors of dyons $$D^{lA}\,,\quad l=1, ... ,\tN\,,\quad A=1, ... , N_f\,.$$
The superpotential in this theory is
\beq
{\mathcal W} = \sqrt{2}\,\sum_{A=1}^{N_f}
\left( \frac{1}{ 2}\,\tilde D_A  b_{U(1)}
D^A +  \tilde D_A  b^p\,T^p  D^A\right)\, 
+{\mathcal W}_{[\mu ]}( b_{U(1)}, b^p),
\label{superpottN}
\eeq
Here $b_{U(1)}$ is a chiral superfield, the \ntwo superpartner of $B^{U(1)}_{\mu}$, 
where $B^{U(1)}_{\mu}$ is a particular linear combination of the dual gauge fields   not interacting
with the $D^J$ dyons. We renormalized $b_{U(1)}$ so that charges of the $D^{lA}$ dyons 
with respect to this field
are  $\frac{1}{2}$.
This amounts to redefining its coupling constant $\tilde{g}^2_{U(1)}$. Moreover, 
$b^p$ is an SU($\tN$) adjoint chiral field, the \ntwo superpartner of the dual SU($\tN$) gauge field.
 Finally, ${\mathcal W}_{[\mu ]}$ is a $\mu$ dependent part of the superpotential, cf. (\ref{msuperpotbr}).

The deformation superpotential ${\mathcal W}_{[\mu ]}$ 
given in Eq. (\ref{msuperpotbr}) can be expressed in terms
of invariants $u_k$, see Eq.~(\ref{u}). Namely, 
\beq
{\mathcal W}_{[\mu ]}= \mu_2\,u_2 -\frac{\mu_2}{N}\,
\left(1-\sqrt{\frac2N}\,\frac{\mu_1}{\mu_2}\right)\,u_1^2\,,
\label{Wmuu}
\eeq 
where $u_2$ and $u_1$ should be understood as functions of $b_{U(1)}$ and  $b^p$. 
These functions are determined by the exact Seiberg--Witten solution. We will treat them
in Sec. \ref{42}.
Note, that with the singlet dyons decoupled, the VEVs of the
non-Abelian dyons are 
\beq
\langle D^{lA}\rangle \! \! = \langle \bar{\tilde{D}}^{lA}\rangle =
\!\!
\frac1{\sqrt{2}}\,\left(
\begin{array}{cccccc}
0 & \ldots & 0 & \sqrt{\xi_{1}} & \ldots & 0\\
\ldots & \ldots & \ldots  & \ldots & \ldots & \ldots\\
0 & \ldots & 0 & 0 & \ldots & \sqrt{\xi_{\tN}}\\
\end{array}
\right),
\label{DvevN1}
\eeq
where the first $\tN$ $\xi$'s are of the order of $\mu m$, see (\ref{roots}).

\subsection{Decoupling the adjoint matter} 
\label{42}

As  will be shown in Sec.~\ref{43}, the masses of the gauge fields and dyons $D^{lA}$ in the U$(\tN$) gauge theory,
 with the superpotential (\ref{superpottN}),  do not exceed $\sqrt{\mu m}$, while 
 the adjoint matter mass of  is of the  order of $\mu$. Therefore, in the limit (\ref{mularge}) the adjoint matter
 decouples. Below scale $\mu$ our theory becomes dual to \none SQCD with the scale
\beq
\tilde{\Lambda}^{N-2\tN}= \frac{\Lambda_{{\mathcal N}=2}^{N-\tN}}{\mu^{\tN}}\,.
\label{tildeL}
\eeq
The only condition we impose to keep this infrared free theory in the weak coupling 
regime is 
\beq
\sqrt{\mu m} \ll \tilde{\Lambda}\,.
\label{wcdual}
\eeq
This means that at large $\mu$ we must keep the quark masses small enough. The 
larger the value of $\mu$ the smaller the quark masses, so that the product $\mu m$
is constrained from above by $\tilde{\Lambda}^2$.
This is always doable. 

We would like to stress that, although this procedure is perfectly justified in the $r=N$ vacuum we work in,
 it does not work, say, in the
 monopole vacua. In these vacua VEVs of the light matter 
(the Abelian monopoles) are of the order of $\sqrt{\mu \Lambda_{{\mathcal N}=2}}$, which, in turn, 
sets 
the mass scale 
in the dual Abelian U(1)$^{N}$ gauge theory \cite{SW1}. Therefore, we cannot decouple 
the adjoint matter keeping the dual theory at weak coupling. As soon as we 
increase $\mu$ well above  the above  scale,  we break the weak coupling 
condition in the dual U(1)$^{N}$ gauge theory.

In contrast, in the $r=N$ vacuum we can take $\mu$ much larger than the 
quark masses and decouple the
adjoint matter. If the condition (\ref{wcdual}) is fulfilled, the dual theory stays at weak
coupling. The reason is that it is the quark masses rather 
than $\Lambda_{{\mathcal N}=2}$ that determine the ``non-Abelian''  roots of 
the Seiberg--Witten curve and VEVs of the non-Abelian dyons, see (\ref{roots}).

Given the  superpotential (\ref{superpottN}) 
we can explicitly integrate out the adjoint matter. To this end we
expand ${\mathcal W}_{[\mu]}$ in powers of $b_{U(1)}$ and  $b^p$,
\beqn
{\mathcal W}_{[\mu ]}( b_{U(1)}, b^p) 
&= & 
c_1 \, \mu_2\, b_{U(1)}^2 + c_2 \,\mu_2\, (b^p)^2
\nonumber\\[3mm]
&+& c_3 \, \mu_2 m \, b_{U(1)} + c_4 \, \mu_2 \,\Lambda_{{\mathcal N}=2}\, b_{U(1)}
\nonumber\\[3mm]
 &+&
O\left(\frac{\mu_2\, (b^p)^4}{\Lambda^2_{{\mathcal N}=2}}\right) + 
O\left(\frac{\mu_2\, b_{U(1)}^3}{\Lambda_{{\mathcal N}=2}}\right),
\label{supexpand}
\eeqn
where
\beq
m=\frac1{N_f}\,\, \sum_{A=1}^{N_f} m_A \,.
\label{m}
\eeq
We then  note that
\beq
c_4=0\, .
\label{c40}
\eeq
Indeed, a nonvanishing $c_4$ would produce a VEV of $b_{U(1)}$ of the order of
 $\Lambda_{{\mathcal N}=2}$ which, in turn, would imply    VEVs of certain 
dyons  $D^{lA}$ to be of the order of $\sqrt{\mu\Lambda_{{\mathcal N}=2}}$, in direct contradiction with Eqs.
(\ref{DvevN1}) and  (\ref{roots}). 

Moreover, since VEVs of $b_{U(1)}$ and $b^p$ are of the order of the
quark masses (rather than $\Lambda_{{\mathcal N}=2}$) we can neglect higher-order
terms in the expansion (\ref{supexpand}) and keep only linear and quadratic 
terms in the $b$ fields. Higher-order terms are suppressed by powers of $m/\Lambda_{{\mathcal N}=2}$.

Now, substituting (\ref{supexpand}) into (\ref{superpottN}) and integrating over
$b_{U(1)}$ and $b^p$ we get the superpotential which depends only on $D^{lA}$. Minimizing it and 
requiring  VEVs of $D^{lA}$ to be given by  (\ref{DvevN1}) (see also (\ref{roots})) we fix the coefficients
$c_1$, $c_2$ and $c_3$,
\beq
c_1= \frac{\tN}{4}\left(1+\gamma\frac{\tN}{N}\right), \qquad c_2=\frac12,\qquad
c_3= \frac{\tN}{\sqrt{2}}\gamma\left(1+\frac{\tN}{N}\right),
\label{cs}
\eeq
where
\beq
\gamma= 1-\sqrt{\frac2N}\,\frac{\mu_1}{\mu_2}.
\label{gamma}
\eeq
After eliminating the adjoint matter the superpotential takes the form
\beqn
{\mathcal W} &=& -\frac1{2\mu_2}\,
\left[ (\tilde{D}_A D^B)(\tilde{D}_B D^A) - \frac{\alpha_{D}}{\tN} (\tilde{D}_A D^A)^2 \right]\, 
\nonumber\\[3mm]
&+& 
\left[m_A-\frac{\gamma\,(1+\frac{\tN}{N})}{1+\gamma\,\frac{\tN}{N}}\,m\right]\,(\tilde{D}_A D^A),
\label{superpotd}
\eeqn
where the color indices are contracted inside each parentheses, while
\beq
\alpha_{D}=\frac{\gamma\,\frac{\tN}{N}}{ 1+\gamma\,\frac{\tN}{N}}\,\,.
\label{alphaD}
\eeq
This equation presents 
our final large-$\mu$ answer for the superpotential of the theory dual to \none SQCD in the (1, ... , $N$) vacuum.
The second term is the dyon mass term while the first one describes the dyon interaction.

One can check that minimization of this superpotential leads to correct dyon VEVs, cf.
Eq.~(\ref{DvevN1}).
Of course, 
the theory with the superpotential (\ref{superpotd}) 
possesses many other vacua in which different dyons (and different number of dyons) develop VEVs.
We consider only one particular vacuum here.  As was explained in
Sec.~\ref{N2duality}, if we  choose the (1, ... , $N$) vacuum in the original theory 
above the  crossover,  then we end up in the $(0, ... ,0, N+1, ... ,N_f)$ vacuum in the dual 
theory below the crossover, see (\ref{jump}). Vacua with different number of condensed $D$'s
seen in (\ref{superpotd}) are spurious. The reason is that if we start from an $r<N$ vacuum in the original theory 
the dual gauge group (below the crossover) would  be different from U$(\tN)$. Thus, the dual theory would not 
be the U$(\tN)$ gauge theory of dyons $D^{lA}$ ($l=1,...,\tN$), with the superpotential (\ref{superpotd}).

Summarizing this section,  we pass to the limit of large $\mu$ decoupling the adjoint matter
in the dual theory. This leaves us with the dual U$(\tN)$ gauge theory 
with the superpotential (\ref{superpotd}). At this point one should ask:
Are we sure that   $\mu$ is large enough
to decouple the adjoint matter in the original theory (\ref{model}), as well as in the dual theory,
so that the original theory  becomes \none 
SQCD? 

Strictly speaking, it is not easy to directly answer this question since in the domain (\ref{wcdual})
the original theory is at strong coupling,  and our control over its dynamics is limited. 
Nevertheless, one can give the following argument. Let us denote
 the low-energy scale of the original \none SQCD as $\Lambda$.  
 In terms of the scale of the original theory  (\ref{model}) at large $\mu$ we have
\beq
\Lambda^{2N-\tN}= \mu^{N}\,\Lambda_{{\mathcal N}=2}^{N-\tN}\,.
\label{Lambda}
\eeq
The (s)quark masses are small,  and the scale of excitations in \none SQCD
is determined by the parameter (\ref{Lambda}). The nonvanishing masses just 
lift the Higgs branch making all vacua isolated. Therefore, if we require that 
\beq
\mu\gg \Lambda
\label{largemuorig}
\eeq
we can be sure that the adjoint mater is decoupled in the original theory. 
Now, the  weak coupling condition for the dual theory (\ref{wcdual}) can be 
rewritten in terms of $\Lambda$ as follows:
\beq
m\ll \Lambda\,\left(\frac{\Lambda}{\mu}\right)^{\frac{3N}{N-2\tN}}\, .
\label{wcdualL}
\eeq
Since the quark mass scale $m$ is at our disposal,  we can  always choose it to be sufficiently small.
Below we assume that both conditions (\ref{largemuorig}) and (\ref{wcdualL}) are met.

If we  further increase $\mu$ (keeping the quark masses fixed) we hit the upper bound in (\ref{wcdualL})
and the dual theory (\ref{superpotd}) goes through a crossover into strong coupling.\footnote{To avoid this, one can 
simultaneously decrease 
$m$.} Still further increase of the parameter $\mu$,  $$\sqrt{\mu m} \gg \Lambda\,,$$ 
brings us in  the weak coupling regime in the original \none SQCD.
In this regime the (s)quark fields   condense thus  completely Higgsing the U$(N)$ gauge group. Non-Abelian strings are formed which confine monopoles. This regime is quite similar to that studied in \cite{EdTo,SYhet,BSYhet} in
the massless version of the theory (\ref{model}),  with a large Fayet--Iliopoulos $D$-term. 

We stress 
that in this domain (large $m$) the (s)quark fields condense, while in our present setup (small $m$) 
the quarks and gauge bosons decay  into the
monopole-antimonopole stringy mesons on CMS.

\subsection{Perturbative mass spectrum}
\label{43}

In this section we briefly discuss the perturbative mass spectrum of the dual U$(\tN)$
gauge theory, with the superpotential (\ref{superpotd}), at large $\mu$.
At first we assume the limit (\ref{masssplit}) for the quark masses.

The U$(\tN)$ gauge group is completely Higgsed, and the masses of the gauge bosons 
are 
\beq
m_{SU(\tN)}=\tilde{g}_2\sqrt{\xi}
\label{Wmassd}
\eeq
for the SU$(\tN)$ gauge bosons, and 
 \beq
m_{U(1)}=\tilde{g}_1\, \sqrt{\frac{N}{2}}\,\sqrt{\xi}\,.
\label{phmassd}
\eeq
for the U(1) gauge boson. Here $\tilde{g}_1$ and $\tilde{g}_2$ are dual gauge couplings
for U(1) and SU$(\tN)$ gauge bosons respectively, while $\xi$ is a common value of  the first
$\tN$ $\xi_P$'s (see Eqs.~(\ref{qxis}) and (\ref{roots})),
\beqn
\xi &=& 2\,\left\{\sqrt{\frac{2}{N}}\,\mu_1\,\hat{m}+\mu_2(m_{\rm last}-\hat{m})\right\},
\qquad m_{\rm last}=m_K,
\nonumber\\[3mm]
K &=& (N+1),...,N_f\,.
\label{xi}
\eeqn
The dyon masses are determined by the $D$-term potential
\beq
 V^{\rm dual}_D =
 \frac{\tilde{g}^2_2}{2}
\left( \bar{D}_A T^p D_A -
\tilde{D}_A T^p \bar{\tilde{D}}^A \right)^2
+ \frac{\tilde{g}^2_1}{8}
\left(|D^A|^2 -|\tilde{D}_A|^2 
\right)^2
\label{Dtermpot}
\eeq
and the $F$-term potential following from the superpotential (\ref{superpotd}). Diagonalizing the quadratic
form given by these two potentials we find that, out of $4\tN N_F$ real degrees of freedom 
of the scalar dyons, $\tN^2$  are eaten by the Higgs mechanism, $(\tN^2-1)$ real scalar dyons have the same mass as the non-Abelian gauge fields, Eq.
(\ref{Wmassd}), while one scalar dyon has mass (\ref{phmassd}). These dyons are scalar superpartners 
of  the SU$(\tN)$  and U(1) gauge bosons in \none massive vector supermultiplets, respectively. 

Another $2(\tN^2-1)$
dyons form a $(1,\tN^2-1)$ representation of the global group (\ref{c+fd}). Their  mass is as follows:
\beq
m_{(1,\tN^2-1)}=\frac{\xi}{\mu_2}=2 \left(m_{\rm last}-\gamma \hat{m}\right)\,,
\label{adjd}
\eeq 
where $\xi$ is given in Eq. (\ref{xi}),
while two real singlet dyons have mass
 \beq
m_{(1,\,1)}= \sqrt{\frac{N}{2}}\,\frac{\xi}{\mu_1}=2 \left(\hat{m} - 
\sqrt{\frac{N}{2}}\,\frac{\mu_2}{\mu_1}\,\Delta m\right).
\label{singld}
\eeq
Masses of $4N\tN$ bifundamental fields  are given by the mass split of $N$ first and $\tN$ last
quark masses, see (\ref{masssplit}),
\beq
m_{(\bar{N},\, \tN)}= \Delta m\,.
\label{bifundd}
\eeq 
All these dyons are the scalar components of the \none chiral multiplets.

We see that the masses of the gauge multiplets and those of chiral matter get a large split in the limit
of  large $\mu$ and small $m_A$. Chiral matter become much lighter than
the gauge multiplets cf. \cite{SYnone,SYrev}.
Most important  is the fact that in the theory (\ref{superpotd}) 
vacuum expectation values  are developed by the light dyons, with masses given by (\ref{singld}) in the limit (\ref{masssplit}). Thus, we have an extreme type-I superconductivity in the vacuum of the dual theory.

For generic quark masses the perturbative excitation spectrum is rather complicated. 
We summarize it here for a particular case
\beq
\sqrt{\frac{\tN}{2}}\,\tilde{g}_1=\tilde{g}_2\,, \qquad \gamma =0\,.
\label{singletr}
\eeq
The first condition means that (with our normalizations)  the gauge couplings 
in the SU$(\tN)$ and U(1) sectors are the same, while the last condition implies that we consider a single-trace deformation superpotential in (\ref{msuperpotbr}).
Under these conditions the masses of the gauge bosons $(A_{\mu})^k_l$ are 
\beq
m_{\rm gauge}=\tilde{g}_2\sqrt{\frac{\xi_k +\xi_l}{2}}\,.
\label{gmassd}
\eeq
Moreover, $\tN^2$ real dyons have the same masses. They are the \none superpartners of the 
massive gauge bosons.
Another $2\tN^2$ real dyons form a $\tN\times\tN$ complex matrix. The masses of the elements of this matrix are 
\beq
m_{KK'} = m_K+m_{K'},\qquad K,K'=(N+1),...,N_f\,.
\label{tN2dyonsmass}
\eeq
The remaining $4N\tN$ of dyons (which become bifundamentals in the limit (\ref{masssplit}))
have masses
\beq
m_{PK} = m_P-m_{K},\qquad P=1,...,N, \qquad K=(N+1),...,N_f\,.
\label{tNNdyonsmass}
\eeq
Again, we see that the dyons with masses (\ref{tN2dyonsmass}) and (\ref{tNNdyonsmass}) are much lighter than
the gauge bosons and their scalar superpartners. It is the diagonal elements of the dyon matrix with the masses (\ref{tN2dyonsmass}) that develop vacuum expectation values.

\section{Strings and confined monopoles at large $\mu$}
\label{largemustr}
\setcounter{equation}{0}

Since in the dual theory (\ref{superpotd}) the dyons develop vacuum expectation values, see Eq. (\ref{Dvev}), 
this theory support strings. Consider the limit (\ref{masssplit}) in which the global color-flavor group 
(\ref{c+fd}) is restored and these strings become non-Abelian. 
As was discussed in Sec.~\ref{43}, the mass terms of those dyons that develop VEVs are much smaller 
than the gauge boson masses in the dual gauge group U$(\tN)$. Therefore, we deal with the type-I superconductor.
A detailed discussion of the non-Abelian string solutions for this case will be presented elsewhere. 
Here we briefly mention certain peculiar features of such strings.

These strings are not BPS-saturated;  their profile functions satisfy second-order equations of motion.
These profile functions have logarithmic long-range tails formed by light dyonic scalars
with masses (\ref{adjd}) and (\ref{singld}), see \cite{Y99} where Abelian strings in the extreme
type-I superconducting vacuum were studied. The string tension in this regime is
\beq
 T = \frac{4\pi |\xi |}{\log{(\tilde{g}\mu/m)}}\,,
\label{tentypeI}
\eeq
while their transverse sizes scale as
\beq
 R \sim \frac{\log{(\tilde{g}\mu/m)}}{\tilde{g}\sqrt{\xi}} \,,
\label{R}
\eeq
with the logarithmic accuracy Here $\xi$ is given in Eq.~(\ref{xi}), and we assume that
$\tilde{g}_2\sim \tilde{g}_1\sim\tilde{g}$.

As was mentioned in Sec.~\ref{stringsN2},
the internal dynamics of the non-Abelian strings in 
the \ntwo limit at small $\mu$ is qualitatively described by an \ntwot supersymmetric 
weighted CP model \cite{HT1,ABEKY,SYmon,HT2}, see also \cite{SYV}. 
In the dual bulk theory, the string world-sheet model is CP$(N_f-1)$
 with $\tN$ positive charges associated with the orientational modes 
 and $N$ negative charges 
 associated with string's size moduli (the latter are specific for semilocal string) 
 \cite{HT1,HT2,SYsem,Jsem,SYdual,SYtorkink}. 

At large $\mu$ the semilocal strings at hand are no longer BPS-saturated. Their
size moduli $\rho_P$ are lifted, and the string tends to shrink in 
type-I superconductors and to expand in type-II superconductors \cite{Hind,AchVas}. 
Remember, we deal with type I. Thus, the shrinkage of the  semilocal strings 
results in conventional local strings. They are stable.  The  size moduli of the  semilocal strings acquire masses 
of the order of 
\beq
m_{\rho}\sim \frac{1}{\tilde{g}\sqrt{\xi}\, R^2}\sim \frac{\tilde{g}\sqrt{\xi}}{\log{(\tilde{g}\mu/m)}}\,,
\label{rhomass}
\eeq
cf. \cite{Hind}.
Then, the world-sheet theory  effectively reduces to CP$(\tN-1)$ model which describes the orientational mode dynamics. In particular, as a matter of fact, the constraint (\ref{unitvec}) is replaced by
\beq
  |n_K|^2 =2\tilde{\beta}\, .
\label{unitvecCP}
\eeq
Another feature of the non-Abelian strings in the extreme type-I superconductors 
is that the coupling constant $\tilde{\beta}$ of the  CP$(\tN-1)$ model  becomes very large,
\beq
\tilde{\beta} \sim \frac{\tilde{g}^2\mu}{m}\,.
\label{betatypeI}
\eeq
This effect is due to the presence of a long-range tail in the string in the type-I superconductor.
Using the one-loop renormalization equation in the asymptotically free CP$(\tN-1)$ model 
\beq
4\pi\tilde{\beta}(\xi) =
\tN \ln{\frac{\sqrt{\xi}}{\Lambda_{CP}}}
\label{d2coupling}
\eeq
we find that the CP$(\tN-1)$ model scale 
becomes exponentially small,
\beq
\Lambda_{\rm CP}\sim {\sqrt{\xi}}\exp{\left(-{\rm const}\,\frac{\tilde{g}^2\mu}{m}\right)} \,.
\label{LambdaCP}
\eeq

\vspace{1mm}

Now, it is time discuss confined monopoles of the bulk theory  corresponding to kinks 
in the world-sheet CP model.
At large $\mu$ the
non-Abelian strings are no longer BPS-saturated, and, consequently,
the \ntwot supersym\-metry in the world-sheet CP 
model is lost. Non-supersymmetric CP$(\tN-1)$ model no longer has $\tN$ degenerate vacua, the true vacuum is unique,
but the model has a family of quasi-vacua \cite{Wtheta,GSY05}. The splittings are of the order of $\Lambda_{CP}$.
Thus, $\tN$ different non-Abelian strings are split in their tensions. This implies 
two-dimensional confinement of monopoles, 
along the string \cite{GSY05}, in addition to their 
permanent attachment to strings. The monopoles cannot move freely along the string. They are combined into
 monopole-antimonopole pairs, the attraction is due to the fact that  the string between the monopole and antimonopole
at hand has a slightly higher tension than the strings outside.

However, this effect is tiny (because of the small value of the
parameter $\Lambda_{\rm CP}$) and does 
{\em not}
 determine the distance between the monopole and antimonopole in the 
 stringy meson in Fig.~\ref{figmeson}. This distance is determined by
the classical string tension (\ref{tentypeI}) itself (and the kink masses),
 rather than the tiny quantum differences between 
 the tensions of different non-Abelian strings. Therefore, we will ignore this effect, the tension splitting.

Another effect which affects the formation of monopole-antimonopole stringy mesons at large $\mu$
is the lifting of the size moduli of the semilocal string, see (\ref{rhomass}). Although the kinks that are in the 
$(1,\tN)$ representation of the global group (\ref{c+fd}) are still light (their masses are of the order 
of $\Lambda_{\rm CP}$), the kinks in the $(N,1)$ representation become heavier. We expect them to have masses
of the order of the masses of the $\rho$-excitations (see (\ref{rhomass})),
\beq
m^{{\rm kink}}_{(N,1)}\sim \frac{\tilde{g}\,\sqrt{\xi}}{\log{(\tilde{g}\mu/m)}}\,.
\label{mkink}
\eeq
 
These kinks (confined monopoles) form stringy mesons in the adjoint 
representation of the SU$(N)$ subgroup of the global group.
We recall that the $(N^2-1,1)$  stringy mesons are 
former (screened) quarks and gauge bosons of the original \none SQCD.
As was already explained, below the crossover (at small $\sqrt{\mu m}$, see (\ref{wcdual})) the quarks and 
gauge bosons decay into the monopole-antimonopole pairs and form stringy mesons $M_P^{P'}$, 
$P=1, ... , N$, shown in Fig.~\ref{figmeson}. 

From the kink mass formulas (\ref{rhomass}) and (\ref{tentypeI})  and 
the string tension we expect the mass of the $M_P^{P'}$ mesons to be
\beq
m_{M_P^{P'}} \sim \sqrt{T}\,,
\label{mstringy}
\eeq
provided the meson spins are of the order of unity.
The masses of these stringy mesons are determined by the string tension, much in the same way as 
in the \ntwo limit, see (\ref{mstringyN2}).

\section{Relation to  Seiberg's duality}
\label{Seiberg}
\setcounter{equation}{0}

The last but not the least
topic to discuss 
is the relation between our duality (and the
monopole confinement mechanism) at large $\mu$ and Seiberg's duality in \none SQCD \cite{Sdual,IS}.
The 
light dyons $D^{lA}$ of our U($\tN$) dual theory  could be  identified with Seiberg's ``dual quarks''. 
This is natural since they carry the same quantum numbers: both are in fundamental representations of the dual gauge group U($\tN$) and  the global flavor group SU$(N_f)$.

Moreover, the  stringy mesons formed by the
monopole-antimonopole pairs correspond to Seiberg's neutral mesons $M_A^B$, $A,B=1,...,N_f$,
which are in the singlet or adjoint representations of global flavor group both in
our and Seiberg's dual descriptions of \none SQCD. This conceptual similarity does not extend further, however.
There is a crucial distinction:  in our dual theory the
stringy mesons are non-perturbative objects and are rather heavy, with masses
determined by the string tension, (\ref{mstringy}). The dual gauge bosons and in particular, dyons $D^{lA}$,
 are much lighter,
see (\ref{Wmassd}), (\ref{phmassd}) and (\ref{adjd}), (\ref{singld}) respectively.

At the same time, in Seiberg's dual theory, the $M_A^B$ mesons
appear as fundamental fields at the Lagrangian level and are light. 
As was already mentioned in Sec.~\ref{intro},
our understanding of these dramatic differences is that Seiberg's duality refers to  
$N$ monopole vacua in  which the meson fields $M_A^B$ condense, making the dyons  (``dual quarks'')
heavy \cite{IS,IS2}. Let us briefly review  how this happens. Consider the U$(\tN)$ version of Seiberg's dual theory with
the superpotential  
\beq
{\mathcal W}_{S} =\sqrt{2}\,
(\tilde{D}_A D^B) M^A_B + \Lambda m_A\, M^A_A\,,
\label{seib}
\eeq
where we conjectured that Seiberg's ``dual quarks'' can be identified with our dyons $D^{lA}$.
Following \cite{IS,IS2}, we assume that the $M_A^B$ fields develop VEVs making dyons heavy and integrate dyons out.
The gluino condensation in the U$(\tN)$ gauge theory with no matter induces the superpotential
\beq
{\mathcal W}_{S}^{\rm eff} = \tN\,\Lambda^{\frac{2\tN -N}{\tN}} \left({\rm det}\,M\right)^{\frac{1}{\tN}} + \Lambda m_A\, M^A_A\,.
\label{seibsup}
\eeq
Strictly speaking the scale of Seiberg's dual theory should appear in the first term here. However,
this scale is estimated to be of the order of the scale of the original \none QCD $\Lambda$ \cite{IS2},
 and 
in this estimate we do not distinguish between the two.

Minimizing this superpotential with respect to $M_A^B$ we find
\beq
\langle M\rangle \sim \Lambda^{\frac{N-\tN }{N}} m^{\frac{\tN}{N}}\,.
\label{MVEV}
\eeq
 The presence of $N$ vacua in \none SQCD is well-known and follows e.g. from Witten's index. 
 It is also known that these vacua
are continuously connected to $N$ monopole vacua of \ntwo SQCD through the
$\mu$
deformation \cite{KSS,EFGR,GVY,CKM}. Since the
``dual quarks''  do not condense in these vacua  the non-asymptotically
free Seiberg's dual  theory is in the  Coulomb phase (``free dyonic phase''). This is true for energies
above the scale of the $M$-field VEVs (\ref{MVEV}). Below this scale, all dyons decouple and  Seiberg's
dual theory becomes  pure Yang-Milles theory with the U($\tN $) gauge group. It flows into the strong coupling, and 
the SU($\tN $) sector becomes confining. The U(1) gauge factor remains unbroken.

\begin{figure}
\epsfxsize=10cm
\centerline{\epsfbox{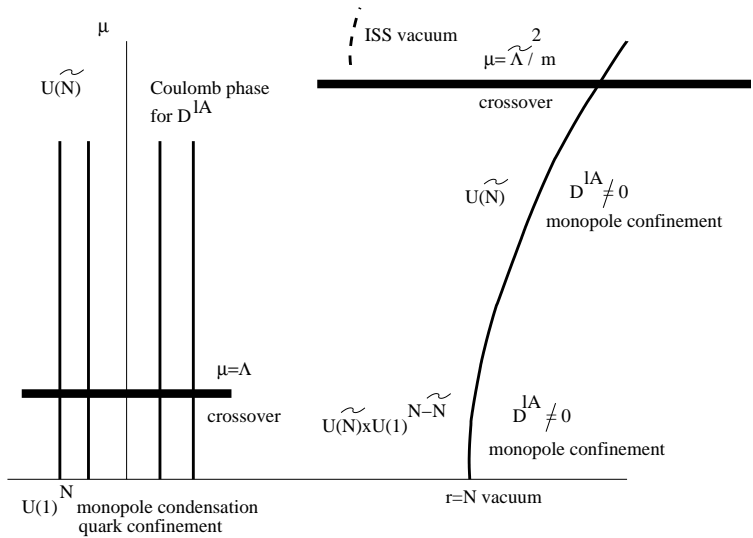}}
\caption{\small $\mu$-evolution of different vacua at small $m$. 
$N$ monopole vacua are shown by thick solid lines
on the left, while $r=N$ vacuum is on the right. The ISS vacuum is shown by thick dashed line.
Gauge groups in different regimes are indicated as well as condensed or confined states.}
\label{figmuevol}
\end{figure}

In Fig.~\ref{figmuevol} we show schematically the evolution of different vacua versus $\mu$ at small $m$.
The vertical axis in this Figure corresponds to  $\mu$, while the horizontal axis schematically represents  
VEVs of various fields in the given vacuum. 
At small $\mu$, near the Coulomb branch of \ntwo SQCD, we have the U(1)$^N$ Abelian gauge 
theories in the $N$ monopole vacua. Condensation of  the monopoles leads to formation of the
electric ANO strings and Abelian confinement of quarks \cite{SW1,SW2}. One U(1) factor remains unbroken.
 At $\mu\sim \Lambda$ these vacua
go through a crossover into the non-Abelian phase. In the limit of infinite $\mu$ they are described 
via Seiberg's dual theory. It is  the U$(\tN)$ infrared-free non-Abelian gauge theory 
with neutral mesonic fields described by the superpotential (\ref{seib}) \cite{Sdual,IS,IS2}. As was
reviewed above, the $M$ fields condense, and the theory is in the Coulomb phase for dyons $D^{lA}$.

Our dual theory applies to the $r=N$ quark vacuum of the original \none SQCD, rather than 
the monopole vacua. In the 
strong coupling regime at small $m$ (described by a weakly coupled  dual theory 
 in the domain (\ref{wcdual})) the
light dyons $D^{lA}$ condense in this vacuum, triggering formation of the
non-Abelian strings with  confinement of monopoles ensuing automatically.
 This vacuum has dyon condensate
proportional to $\sqrt{\mu m}$, see (\ref{DvevN1}),  and represents a run-away vacuum not seen in Seiberg's
dual description, where $\mu$ is considered to be strictly  infinite.

There exists a bunch of other ``hybrid" vacua in the theory, in which   
 at small $\mu$ we deal with $r<N$ quarks and some monopoles condensing.
All of them have unbroken U(1) gauge group \cite{Cachazo2}. We do not study them in this paper.

A few words about  the relation of our $r=N$ vacuum to the Intriligator--Seiberg--Shih  vacuum \cite{ISS}.
This vacuum looks rather similar to ours. The dyons $D^{lA}$ (Seiberg's ``dual quarks'') condense in both
of these vacua. However, clearly these vacua are different. In particular, for a generic choice of 
the quark masses,
 supersymmetry is broken in the ISS vacuum, while the
 $r=N$ vacuum is supersymmetric. Also, the ISS vacuum has dyon VEVs
of the order of $\sqrt{m\Lambda}$ while in 
the $r=N$ vacuum they are much larger, proportional to $\sqrt{\mu m}$. Still, the presence
of the $D^{lA}$ condensate indicates that the ISS vacuum could have physics  similar to that in 
our $r=N$ vacuum. In 
particular, it could exhibit confinement of monopoles and a phenomenon similar to our 
decay of quarks and gauge bosons of the original \none SQCD
into the monopole-antimonopole stringy mesons.
Since the ISS vacuum
is not supersymmetric,  it may not exist at all $\mu$. This would explain why we do not see this vacuum
in our dual  theory (\ref{superpotd}) in the domain (\ref{wcdualL}).
We show this vacuum by the dashed line in Fig.~\ref{figmuevol}. 
The fate of the ISS vacuum in the framework of our construction calls for further studies.

\section{Conclusions}
\label{conclu}

Let us summarize our findings. We started from our recent development of the
 non-Abelian duality in the {\em quark vacua} of \ntwo super-Yang--Mills 
theory with the U$(N)$ gauge group and $N_f$ flavors ($N_f>N$). The fact that $N_f>N$ is very crucial, as
will be emphasized below. The quark mass terms are introduced in a judiciously chosen way.
Instead of the Fayet--Iliopoulos term of the $D$ type, as previously, we introduce it through a superpotential (i.e. $F$ type).
We construct dual pairs.
Both theories from the dual 
pair support non-Abelian strings which confine monopoles. 

Next we undertake a next step, basically  in the uncharted waters. we introduce an \ntwo-breaking deformation, 
a mass term $\mu{\mathcal A}^2$
for the adjoint fields. Our final goal is to make the adjoint fields heavy and thus pass to \none SQCD.

Starting from a small deformation we eventually make it large which enforces 
complete decoupling of the adjoint fields. We show that the \ntwo non-Abelian duality fully survives in 
the limit of \none SQCD, albeit some technicalities change. For instance, non-Abelian strings which used 
to be BPS-saturated in the \ntwo limit, cease to be saturated in \none SQCD. 
They become strings typical of the extreme type-I superconducting regime.

Our 
duality is a distant relative of Seiberg's duality in \none SQCD. Both share  common features 
but have many drastic distinctions. This is due to the fact that Seiberg's duality apply to the monopole
rather than quark vacua.

More specifically, in our theory we deal with   $N< N_f<\frac32 N $ massive quark flavors. 
 We  consider the vacuum in which  $N$ squarks condense.
Then we identify a crossover transition 
from weak to strong coupling. At strong coupling we find a dual theory, U$(N_f-N)$ SQCD, with   
$N_f$ light dyon flavors. The dual theory is at weak coupling provided $\mu m$ is small enough
(at large $\mu$ this requires taking $m$ to be rather  small).
 Condensation of light dyons
$D^{lA}$ in this theory triggers the formation of non-Abelian strings and confinement of monopoles.
Quarks and gauge bosons of the original \none SQCD decay into the monopole-antimonopole pairs on CMS and 
form stringy mesons shown in Fig.~\ref{figmeson}.

We would like to stress that the condition $\tN>1$ is crucial for our construction.
As was explained in Sec.~\ref{largemu}, the presence of the dual non-Abelian group allows us to increase $\mu$,
eventually
decoupling the adjoint field and, simultaneously, keeping the dual theory at weak coupling. The reason is that
we can take quark masses rather small to satisfy the weak coupling condition (\ref{wcdual}). If the dual
gauge group were Abelian, the light matter  VEV's would be of the order of $\sqrt{\mu\Lambda_{{\cal N}=2}}$, hence
the theory would go into the strong coupling regime once we increase $\mu$ above
$\Lambda_{{\cal N}=2}$.

\section*{Acknowledgments}
 The work of MS was supported in part by DOE
grant DE-FG02-94ER408. 
The work of AY was  supported
by  FTPI, University of Minnesota,
by RFBR Grant No. 09-02-00457a
and by Russian State Grant for
Scientific Schools RSGSS-65751.2010.2.

\section*{Appendix:  \\
U(3) theory with \boldmath{$N_f=5$} at small \boldmath{$\mu$}}

 \renewcommand{\theequation}{A.\arabic{equation}}
\setcounter{equation}{0}
 
 \renewcommand{\thesubsection}{A.\arabic{subsection}}
\setcounter{subsection}{0}

In this Appendix 
  following \cite{SYdual}  we  consider specific example of U(3) gauge theory
with $N_f=5$ quark flavors (so that $N=3$, $\tN=2$) and present the low-energy dual
theory at small values of  FI parameter, see (\ref{strcoup}). The  gauge group (\ref{dualgaugegroup})
in this case has the form
\beq
{\rm U}(2)\times {\rm U}(1)_8\times {\rm U}(1)\,,
\label{U2dualgaugegroup}
\eeq
where U$(1)_8$ denotes a U(1) factor of the gauge group which is associated with
$T_8$ generator of the U(3) gauge group of the original theory. 

The bosonic part of the effective low-energy
action of the theory in the domain (\ref{strcoup}) has the form
\beqn
 S_{{\rm dual}} &=&\int d^4x \left[\frac{1}{4\tilde{g}^2_{2}}
\left(F^{p}_{\mu\nu}\right)^2 
+\frac1{4g^2_1}\left(F_{\mu\nu}\right)^2 +\frac1{4\tilde{g}^2_8}\left(F^8_{\mu\nu}\right)^2
+\frac1{\tilde{g}^2_2}\left|\pt_{\mu}b^p\right|^2 
\right.
\nonumber\\[4mm]
&+& \frac1{g^2_1}
\left|\partial_{\mu}a\right|^2 +\frac1{\tilde{g}^2_8}
\left|\partial_{\mu}b^8\right|^2
+\left|\nabla^1_{\mu}
D^A\right|^2 + \left|\nabla^1_{\mu} \tilde{D}_A\right|^2+
\left|\nabla^2_{\mu}
D^3\right|^2 + \left|\nabla^2_{\mu} \tilde{D}_{3}\right|^2
\nonumber\\[4mm]
&+&
\left. V(D,\tilde{D},b^p,b^8,a)\right]\,,
\label{SIII}
\eeqn

Here  $B^{p}_{\mu}$ ($p=1,2,3$), $B^{8}_{\mu}$ and $A_{\mu}$ are gauge fields of (\ref{U2dualgaugegroup}),
while $F_{\mu\nu}^p$, $F_{\mu\nu}^8$ and $F_{\mu\nu}$ are  their field strengths.
Their scalar \ntwo superpartners
 $b^p$ and  $b^8$ in terms of the fields of the original theory (\ref{model})
have  the form
\beq
b^3= \frac{1}{\sqrt{2}}\,(a^{3}+a^{3}_D)\;\;\; {\rm for}\;\;\; p=3,
\qquad b^8= \frac{1}{\sqrt{10}}\,(a^{8}+3a^{8}_D),
\label{bbp}
\eeq
where subscript $D$ means dual scalar fields \cite{SW1,SW2}, 
while field $a$ is the same as in (\ref{model}).
Covariant derivatives are defined in accordance
with the charges of the  $D^{lA}$ and $D^3$ dyons, see \cite{SYdual} for more details. Namely,
\beqn
\nabla^1_\mu & = & 
=\pt_{\mu}-i\left(\frac12 A_{\mu}+\sqrt{2}\,B^p_{\mu}\frac{\tau^p}{2}
 +\frac12\sqrt{\frac{10}{3}}\,B^8_{\mu}\right)\,,
\nonumber\\[3mm]
\nabla^2_\mu & = & 
=\pt_{\mu}-i\left(\frac12 A_{\mu}
 -\sqrt{\frac{10}{3}}\,B^8_{\mu}\right)\,.
\label{nablaD}
\eeqn
The coupling constants $g_1$, $\tilde{g}_8$ and $\tilde{g}_2$ 
correspond to two U(1) and the SU(2) gauge groups, respectively.
The scalar potential $V(D,\tilde{D},b^p,b^8,a)$ in the action (\ref{SIII})
is 
\beqn
&& V(D,\tilde{D},b^p,b^8,a) =
 \frac{\tilde{g}^2_2}{4}
\left( \bar{D}_A\tau^p D_A -
\tilde{D}_A \tau^p \bar{\tilde{D}}^A \right)^2
\nonumber\\[3mm]
&+& \frac{10}{3}\frac{\tilde{g}^2_8}{8}
\left(|D^A|^2 -|\tilde{D}_A|^2 -2|D^3|^2 +
2|\tilde{D}_3|^2 \right)^2
\nonumber\\[3mm]
&+& \frac{\tilde{g}^2_1}{8}
\left(|D^A|^2 -|\tilde{D}_A|^2 +|D^3|^2 -
|\tilde{D}_3|^2 
\right)^2
\nonumber\\[3mm]
&+& \frac{\tilde{g}_2^2}{2}
\left| \sqrt{2}\tilde{D}_A \tau^p D_A +\sqrt{2}\,\,\frac{\pt{\mathcal W_{\mu}}}{\pt b^p}
\right|^2+
\frac{\tilde{g}^2_1}{2}\left| \tilde{D}_A D^A+
\tilde{D}_3 D_3 +\sqrt{2}\,\,\frac{\pt{\mathcal W}_{\mu}}{\pt a}\right|^2
\nonumber\\[3mm]
&+& 
\frac{\tilde{g}_8^2}{2}\left| \sqrt{\frac{10}{3}}\tilde{D}_A D^A-
2\sqrt{\frac{10}{3}}\tilde{D}_3 D^3 + \sqrt{2}\,\,\frac{\pt{\mathcal W}_{\mu}}{\pt b^8}\right|^2
\nonumber\\[3mm]
&+&\frac12 \left\{ \left|(a+\tau^p\sqrt{2}\,b^p +\sqrt{\frac{10}{3}}\,b^8+\sqrt{2}m_A
)D^A\right|^2 
\right.
\nonumber\\[3mm]
&+& 
\left|(a+\tau^p\sqrt{2}\,b^p +\sqrt{\frac{10}{3}}\,b^8+\sqrt{2}m_A
)\bar{\tilde{D}}_A\right|^2
\nonumber\\[3mm]
&+&\left.
\left|\;a-2\sqrt{\frac{10}{3}}\,b^8+\sqrt{2}m_3 \;
\right|^2\left(|D^3|^2+|\tilde{D}_3|^2\right) \right\}\,.
\label{potIII}
\eeqn

\vspace{2mm}

The theory (\ref{SIII}) is at weak coupling in the domain (\ref{strcoup}) but
the derivatives of the superpotential (\ref{msuperpotbr}) entering in (\ref{potIII})
(which determine VEVs of dyons)
are rather complicated functions of fields $a$, $b^8$ and $b^p$. In \cite{SYfstr} 
we used the exact
Seiberg-Witten solution of our theory to determine these derivatives in $r=N$
vacuum. Here we briefly review this calculation.

First we make a quantum generalization
\beq
\frac{\pt{\mathcal W}_{\mu}}{\pt b^p}\to\mu_2 \,\frac{\pt u_2}{\pt b^p}\,,
\qquad \frac{\pt{\mathcal W}_{\mu}}{\pt b^8}\to\mu_2 \,\frac{\pt u_2}{\pt b^8}\,,
\qquad \frac{\pt{\mathcal W}_{\mu}}{\pt a}\to\mu_1 \sqrt{\frac{2}{N}}\,\frac{\pt u_2}{\pt a}\,,
\label{dwda}
\eeq
where 
\beq
u_k= \bra {\rm Tr}\left(\frac12\, a + T^a\, a^a\right)^k\ket, \qquad k=1, ..., N\,,
\label{u}
\eeq
are gauge invariant parameters which describe  the Coulomb branch.

To select the desired vacuum (1,2,3) ( which transforms into (4,5,3) vacuum below crossover) among all other vacua in the Seiberg-Witten
curve we require that the curve has the factorized form (\ref{rNcurve}), while
the double roots $e_P$ are semiclassically (at large masses) are given  by mass
parameters, $\sqrt{2}e_P\approx -m_P$, $P=1,...,N$.

Using explicit expressions from \cite{ArFa,KLTY,ArPlSh,HaOz} which express derivatives
of $u_k$ with respect to scalar fields $a^a$ ($a=1,2,3$) of the original theory 
(\ref{model}) and taking
into account monodromies which convert these derivatives into derivatives with respect to
$b^p$, $b^8$ and $a$ \cite{SYfstr,SYdual} we obtain
\beqn
\frac{\pt u_2}{\pt a}
&=&
e_1+e_2+e_3\,, \qquad \frac1{\sqrt{2}}\frac{\pt u_2}{\pt b^3}=e_1-e_2\,,
\nonumber\\[3mm]
 \frac1{\sqrt{10}}\frac{\pt u_2}{\pt b^8}
 &=&
 \frac1{\sqrt{3}}(e_1+e_2-2e_3)\,,
\label{duda32}
\eeqn
where $e_P$ are double roots of the Seiberg-Witten curve (\ref{curve}) with  shifted
masses (\ref{shift}) for the case $N=3$, $\tN=2$.

Vacua of the theory (\ref{SIII}) are determined by zeros of all $D$ and $F$-terms in 
(\ref{potIII}). Using the  derivatives of the superpotential (\ref{msuperpotbr})
obtained above we get the VEV's of dyons in the form
\beqn
\langle D^{lA}\rangle &=& \langle \bar{\tilde{D}}^{lA}\rangle = 
\frac1{\sqrt{2}}\,
\left(
\begin{array}{ccccc}
0 &  0 & 0 & \sqrt{\xi_1} & 0\\
0 &  0 & 0 & 0 & \sqrt{\xi_2}\\
\end{array}
\right),
\nonumber\\[4mm]
\langle D^{3}\rangle &=& \langle\bar{\tilde{D}}^{3}\rangle = \sqrt{\frac{\xi_3}{2}}, 
\label{Dvev32}
\eeqn
where FI parameters $\xi_P$ are determined by (\ref{qxis}). The obvious 
generalization of this formula to an arbitrary $N$ and $\tN$ gives Eq.~(\ref{Dvev}) quoted in the main text.


\newpage 
\small
\begin{thebibliography}{99}
\addcontentsline{toc}{section}{References}
\itemsep -2pt

\bibitem{mandelstam}
Y.~Nambu,
  Phys.\ Rev.\  D {\bf 10}, 4262 (1974);\\
G.~'t Hooft,
{\em Gauge theories with unified weak, electromagnetic and strong interactions,}
in Proc. of the E.P.S. Int. Conf. on High Energy Physics, Palermo, 23-28 June, 1975
ed. A. Zichichi (Editrice Compositori, Bologna, 1976);
Nucl.\ Phys.\ B {\bf 190}, 455 (1981);
S.~Mandelstam,
Phys.\ Rept.\  {\bf 23}, 245 (1976).

\bibitem{SW1}
N.~Seiberg and E.~Witten,
Nucl. Phys. {\bf B426}, 19 (1994),
(E) {\bf B430},  485 (1994) [hep-th/9407087].

 \bibitem{SW2}
N.~Seiberg and E.~Witten,
Nucl. Phys. {\bf B431}, 484  (1994)
[hep-th/9408099].

\bibitem{ANO}
A.~Abrikosov, Sov.~Phys. JETP {\bf32}, 1442  (1957)
[Reprinted in {\em Solitons and Particles}, Eds. C. Rebbi and G. Soliani
(World Scientific, Singapore, 1984), p. 356];
H.~Nielsen and P.~Olesen, Nucl.~Phys. {\bf B61}, 45 (1973)
[Reprinted in {\em Solitons and Particles}, Eds. C. Rebbi and G. Soliani
(World Scientific, Singapore, 1984), p. 365].

\bibitem{DS}
M.~R.~Douglas and S.~H.~Shenker,
Nucl.\ Phys.\ B {\bf 447}, 271 (1995)
[hep-th/9503163].

\bibitem{HSZ}
A.~Hanany, M.~J.~Strassler and A.~Zaffaroni,
Nucl.\ Phys.\ B {\bf 513}, 87 (1998)
[hep-th/9707244].

\bibitem{Strassler}
M.~Strassler,
  Prog.\ Theor.\ Phys.\ Suppl.\  {\bf 131}, 439 (1998)
  [hep-lat/9803009].

\bibitem{VY}
A.~I.~Vainshtein and A.~Yung,
Nucl.\ Phys.\ B {\bf 614}, 3 (2001)
[hep-th/0012250].

\bibitem{Yrev}
A.~Yung,
{\em What Do We Learn About Confinement From The Seiberg--Witten Theory?},
Proc. of 28th PNPI Winter School of Physics, St. Petersburg, Russia, 2000,
[hep-th/0005088];  published in  {\sl At the frontier of particle physics}, Ed. M.~Shifman,
(World Scientific, Singapore, 2001)
vol. 3, p. 1827.

\bibitem{SYdual}
  M.~Shifman and A.~Yung,
  Phys.\ Rev.\  D {\bf 79}, 125012 (2009)
  [arXiv:0904.1035 [hep-th]].

\bibitem{SYtorkink}
M.~Shifman and A.~Yung,
  Phys.\ Rev.\  D {\bf 81}, 085009 (2010)
  [arXiv:1002.0322 [hep-th]].

\bibitem{FI}
  P.~Fayet and J.~Iliopoulos,
  Phys.\ Lett.\  B {\bf 51}, 461 (1974).

\bibitem{HT1}
A.~Hanany and D.~Tong,
JHEP {\bf 0307}, 037 (2003)
[hep-th/0306150].

\bibitem{ABEKY}
R.~Auzzi, S.~Bolognesi, J.~Evslin, K.~Konishi and A.~Yung,
Nucl.\ Phys.\ B {\bf 673}, 187 (2003)
[hep-th/0307287].

 \bibitem{SYmon}
M.~Shifman and A.~Yung,
Phys.\ Rev.\ D {\bf 70}, 045004 (2004)
[hep-th/0403149].

\bibitem{HT2}
A.~Hanany and D.~Tong,
JHEP {\bf 0404}, 066 (2004)
[hep-th/0403158].
  
 \bibitem{Trev}
D.~Tong, {\em TASI Lectures on Solitons,}
  arXiv:hep-th/0509216.

\bibitem{Jrev}
  M.~Eto, Y.~Isozumi, M.~Nitta, K.~Ohashi and N.~Sakai,
  J.\ Phys.\ A  {\bf 39}, R315 (2006)
  [arXiv:hep-th/0602170].
  
  \bibitem{SYrev}
M.~Shifman and A.~Yung,
{\sl Supersymmetric Solitons,}
Rev.\ Mod.\ Phys. {\bf 79} 1139 (2007)
[arXiv:hep-th/0703267]; an expanded version in Cambridge University Press, 2009.

\bibitem{Trev2}
D.~Tong,
  Annals Phys.\  {\bf 324}, 30 (2009)
  [arXiv:0809.5060 [hep-th]].

  \bibitem{SYcross}
M.~Shifman and A.~Yung,
Phys. Rev. {\bf D 79}, 105006 (2009)
  arXiv:0901.4144 [hep-th].

\bibitem{SYcrossp}
M.~Shifman and A.~Yung,
  AIP Conf.\ Proc.\  {\bf 1200}, 194 (2010)
  [arXiv:0910.3007 [hep-th]].
  
  \bibitem{APS}
P.~Argyres, M.~Plesser and N.~Seiberg,
Nucl. Phys. {\bf B471}, 159  (1996)
[hep-th/9603042].

 \bibitem{Sdual}
  N.~Seiberg,
  Nucl.\ Phys.\  B {\bf 435}, 129 (1995)
  [arXiv:hep-th/9411149].
    
\bibitem{IS}
K.~A.~Intriligator and N.~Seiberg,
  Nucl.\ Phys.\ Proc.\ Suppl.\  {\bf 45BC}, 1 (1996)
  [hep-th/9509066].

\bibitem{We}
E.~J.~Weinberg,
Nucl.\ Phys.\ B {\bf 167}, 500 (1980);
Nucl.\ Phys.\ B {\bf 203}, 445 (1982).

\bibitem{ISS}
K.~Intriligator, N.~Seiberg and D.~Shih,
  JHEP {\bf 0604}, 021 (2006)
  [hep-th/0602239].
  
\bibitem{IS2}
K.~A.~Intriligator and N.~Seiberg,
Class. \ Quant. \ Grav. \ {\bf 24}, S741-S772 (2007)
[arXiv:hep-ph/0702069].

\bibitem{Shifman:2007kd}
  M.~Shifman and A.~Yung,
  Phys.\ Rev.\  D {\bf 76}, 045005 (2007)
  [arXiv:0705.3811 [hep-th]].
  
\bibitem{komar}
Z.~Komargodski,
  JHEP {\bf 1102}, 019 (2011)
  [arXiv:1010.4105 [hep-th]].

\bibitem{SYfstr}
M.~Shifman and A.~Yung,
  Phys.\ Rev.\  D {\bf 82}, 066006 (2010)
  [arXiv:1005.5264 [hep-th]].
  
  \bibitem{CKM}
G.~Carlino, K.~Konishi and H.~Murayama,
Nucl.\ Phys.\ B {\bf 590}, 37 (2000)
[hep-th/0005076].

\bibitem{BF}
A.~Bilal and F.~Ferrari,
  Nucl.\ Phys.\  B {\bf 516}, 175 (1998)
  [arXiv:hep-th/9706145].

\bibitem{AchVas}
A.~Achucarro and T.~Vachaspati,
  Phys.\ Rept.\  {\bf 327}, 347 (2000)
  [hep-ph/9904229].

\bibitem{SYsem}
 M.~Shifman and A.~Yung,
  Phys.\ Rev.\  D {\bf 73}, 125012 (2006)
  [arXiv:hep-th/0603134].

\bibitem{Jsem}
M.~Eto, J.~Evslin, K.~Konishi, G.~Marmorini, M.~Nitta, K.~Ohashi, W.~Vinci, N.~Yokoi,
  Phys.\ Rev.\  D {\bf 76}, 105002 (2007)
  [arXiv:0704.2218 [hep-th]].
  
\bibitem{SYV}
M. Shifman, W. Vinci, and A. Yung,   in preparation.

\bibitem{T}
D.~Tong,
Phys.\ Rev.\ D {\bf 69}, 065003 (2004)
[hep-th/0307302].

\bibitem{W79}
E.~Witten,
Nucl.\ Phys.\ B {\bf 149}, 285 (1979).

 \bibitem{HoVa}
  K.~Hori and C.~Vafa,
{\em Mirror symmetry,}
  arXiv:hep-th/0002222.
  
  \bibitem{EdTo}
 M.~Edalati and D.~Tong,
  JHEP {\bf 0705}, 005 (2007)
  [arXiv:hep-th/0703045].
  
\bibitem{SYhet}
  M.~Shifman and A.~Yung,
  Phys.\ Rev.\  D {\bf 77}, 125016 (2008)
  [arXiv:0803.0158 [hep-th]].
  
  \bibitem{BSYhet}
P.~A.~Bolokhov, M.~Shifman and A.~Yung,
  Phys. \ Rev. \ D {\bf 79}, 085015 (2009) (Erratum: Phys. Rev. D 80, 049902 (2009))
  [arXiv:0901.4603 [hep-th]].

\bibitem{SYnone}
  M.~Shifman and A.~Yung,
  Phys.\ Rev.\ D {\bf 72}, 085017 (2005) [hep-th/0501211].

\bibitem{Y99}
  A.~Yung,
  Nucl.\ Phys.\ B {\bf 562}, 191 (1999)
  [hep-th/9906243].
  
  \bibitem{Hind}
  M.~Hindmarsh,
  Nucl.\ Phys.\  B {\bf 392}, 461 (1993)
  [arXiv:hep-ph/9206229].
 
\bibitem{Wtheta}
E.~Witten,
Phys.\ Rev.\ Lett.\  {\bf 81}, 2862 (1998)
[hep-th/9807109].

\bibitem{GSY05}
  A.~Gorsky, M.~Shifman and A.~Yung,
  Phys.\ Rev.\  D {\bf 71}, 045010 (2005)
  [arXiv:hep-th/0412082].
  
   \bibitem{KSS}
D.~Kutasov, A.~Schwimmer, and N.~Seiberg,
Nucl. \ Phys.  \ {\bf B459},  455 (1996)

\bibitem{EFGR}
S.~Elitzur, A.~Forge, A.~Giveon and  E.~Rabinovici,
Phys. \ Lett. \ {\bf B353},  79 (1995)
  [arXiv:hep-th/9504080]

\bibitem{GVY}
A.~Gorsky, A.~Vainshtein and  A.~Yung,
Nucl. \ Phys.  \ {\bf B584},  197 (2000)
[arXiv:hep-th/0004087]

\bibitem{Cachazo2}
F.~Cachazo, N.~Seiberg and E.~Witten, 
 	JHEP {\bf 0304}, 018 (2003)   [arXiv:hep-th/0303207] 

\bibitem{ArFa}
P. C.~Argyres and A. E.~Faraggi,
Phys. \ Rev. \ Lett. {\bf 74},  3931 (1995)
[hep-th/9411057].

\bibitem{KLTY}
A.~Klemm, W.~Lerche, S.~Yankielowicz and S.~Theisen,
Phys. \ Lett. \ B {\bf 344}, 169 (1995) 
[hep-th/9411048].

\bibitem{ArPlSh}
P. C. Argyres, M. R. Plesser, and  A. Shapere,
Phys. \ Rev. \ Lett.  \ {\bf 75}, 1699 (1995)
[hep-th/9505100].

\bibitem{HaOz}
A.~Hanany and  Y.~Oz,
Nucl. \ Phys. \ B {\bf 452}, 283 (1995)
[hep-th/9505075].




\end{thebibliography}
\end{document}